\documentclass[aps, prc, twocolumn, superscriptaddress,floatfix, amsmath,amssymb]{revtex4}
\usepackage{color}
\usepackage{graphicx}
\usepackage{CJK}
\usepackage{longtable,booktabs}
\usepackage{ltablex}
\usepackage{bm}
\usepackage{subfigure}
\usepackage{float}
\usepackage[
            pdfstartview=FitH,
            CJKbookmarks=true,
            bookmarksnumbered=true,
            bookmarksopen=true,
            colorlinks=false,
            pdfborder=001,
            linkcolor=blue,
            anchorcolor=blue,
            citecolor=blue
            ]{hyperref}

\newcommand{\blue}[1]{\textcolor{blue}{{#1}}}
\newcommand{\beq}{\begin{equation}}
\newcommand{\eeq}{\end{equation}}

\begin{document}
\begin{CJK*}{GBK}{song}
\title{Influence of triaxial deformation on wobbling motion in even-even nuclei}

\author{Bin Qi} \email{bqi@sdu.edu.cn}
\affiliation{Shandong Provincial Key Laboratory of Optical Astronomy
and Solar-Terrestrial Environment, School of Space Science and
Physics, Institute of Space Sciences, Shandong University, Weihai,
264209, People's Republic of China}

\author{Hui Zhang}
\affiliation{Shandong Provincial Key Laboratory of Optical Astronomy
and Solar-Terrestrial Environment, School of Space Science and
Physics, Institute of Space Sciences, Shandong University, Weihai,
264209, People's Republic of China}

\author{Shou Yu Wang}
\affiliation{Shandong Provincial Key Laboratory of Optical Astronomy
and Solar-Terrestrial Environment, School of Space Science and
Physics, Institute of Space Sciences, Shandong University, Weihai,
264209, People's Republic of China}

\author{Qi Bo Chen}\email{qbchen@pku.edu.cn}
\affiliation{Physik-Department, Technische Universit\"{a}t
M\"{u}nchen, D-85747 Garching, Germany}

\begin{abstract}
The influence of triaxial deformation $\gamma$ on the purely
collective form of wobbling motion in even-even nuclei are
discussed based on the triaxial rotor model. It is found
that the harmonic approximation is realized well when
$\gamma=30^{\circ}$ for the properties of energy spectra and
electric quadrupole transition probabilities, while this
approximation gets bad when $\gamma$ deviates from
$30^{\circ}$. A recent data from Coulomb excitation
experiment, namely $3_1^+$ and $2_2^+$ for the $^{110}$Ru
are studied and might be suggested as the bandhead of the
wobbling bands. In addition, two types of angular momentum
geometries for wobbling motion, stemming from different
$\gamma$ values, are exhibited by azimuthal plots.

\end{abstract}

\maketitle

\section{Introduction}

Two unambiguous fingerprints of the stable triaxiality of nuclei are
chirality~\cite{Meng97} and wobbling~\cite{Bohr75}, which have been
studied actively over the past two decades. Wobbling motion was
introduced by Bohr and Mottelson in 1970s~\cite{Bohr75}. It is
described as small amplitude oscillation of the total angular
momentum vector with respect to the principal axis with the
largest moment of inertia. Since 2001, wobbling experimental
evidence was first reported in $^{163}$Lu~\cite{Lu1631,Lu1632},
and later in $^{161}$Lu, $^{165}$Lu, $^{167}$Lu, $^{167}$Ta
nuclei~\cite{Lu161,Lu165,Lu167,Ta167}. In recent years,
wobbling was reported in other regions as well: $^{135}$Pr,
$^{133}$La in the $A\sim$130 region~\cite{Pr135, Sensharma2019PLB, La133},
$^{105}$Pd in the $A\sim$100 region~\cite{Pd105}, $^{187}$Au and $^{183}$Au in the
$A\sim$190 region~\cite{Au187, Au183}. It is interesting to note that all
of the aforementioned wobbling motions are in odd-$A$ nuclei.

For odd-$A$ nuclei, Frauendorf and D\"{o}nau showed two different
possibilities of wobbling modes: longitudinal case and transverse
case~\cite{Frauendorf14}. The theoretical descriptions of wobbling motion
of odd-$A$ nucleus have been attracted great attention, and extensively
studied with the triaxial particle rotor model (PRM)~\cite{Hamamoto02,
Hamamoto2003PRC, Hagemann05, Frauendorf14, Streck18} and
its approximation solutions~\cite{Tanabe17, Budaca2018PRC, Raduta2020PRC},
the random phase approximation~\cite{Matsuzaki02, Matsuzaki2003PRC, Matsuzaki2004PRC,
Matsuzaki2004PRC_v1, Shimizu2005PRC, Shimizu2008PRC, Frauendorf2015PRC, Nakatsukasa16},
the angular momentum projection (AMP) methods~\cite{Shimada18, Sensharma2019PLB},
or the collective Hamiltonian method~\cite{CHENQB14, CHENQB16}. There are
also some debates on the interpretations~\cite{Frauendorf14} for the
wobbling in odd-$A$ nucleus~\cite{Tanabe17, Frauendorf182, Tanabe18, Lawrie20}.

Meanwhile, wobbling modes in even-even nuclei~\cite{Bohr75} has been
studied continuously in theory, e.g., see Refs.~\cite{Marshalek79,
Mikhailov78, Shimizu95, Oi00, Casten03, Oi07,CQB15,Raduta07}.
Recently, two new bands built on the two-quasiparticle
$\pi(h_{11/2})^{2}$ configuration were reported in even-even nuclei
$^{130}$Ba~\cite{Petrache2019PLB}, which were lately interpreted as
the transverse wobbling bands by PRM~\cite{Ba130} and AMP
method~\cite{Zhao20}. However, one notes that the experimental
evidence for the wobbling motion based on even-even nucleus with
zero quasi-particle configuration, namely the originally predicted
purely collective form~\cite{Bohr75}, is fragmentary yet. For
instance, the possible evidence was pointed to the $\gamma$-band in
$^{112}$Ru~\cite{Ru112}. Unfortunately, there were not interband
$\gamma$ rays connecting between the candidates of wobbling band.

The recent advent of new-generation detectors has been opening a
great possibility to explore a new area of the collective rotation
physics, in which one interesting exploration is searching for the
wobbling mode with purely collective form. Prior to this, the
investigation for the variation of the wobbling excited bands with respect
to the triaxial parameter $\gamma$ could be helpful for the experimental
exploration. In addition, a clear picture of the angular momentum geometry
and its evolution for the wobbling excitation with purely collective
form will also be helpful to better understand the wobbling
phenomena in odd-$A$ nuclei. Motivated by the above considerations,
in this paper we discuss systematically the wobbling excitation
in even-even nuclei using triaxial rotor model.

\section{Discussion}

\subsection{Influence of $\gamma$ value on the harmonic approximation}

\begin{figure}[ht!]
\includegraphics[width=8cm]{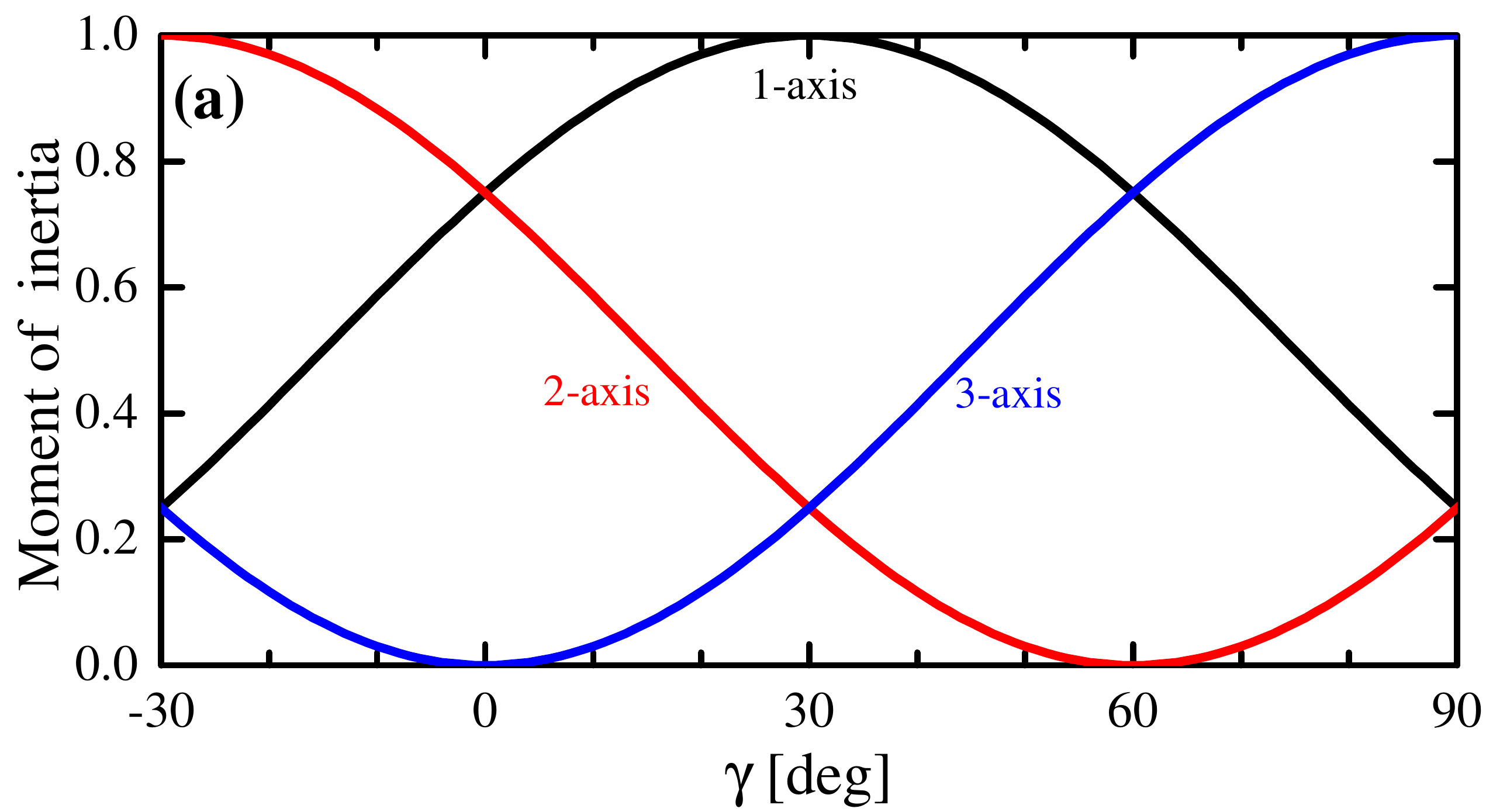}
\includegraphics[width=8.1cm]{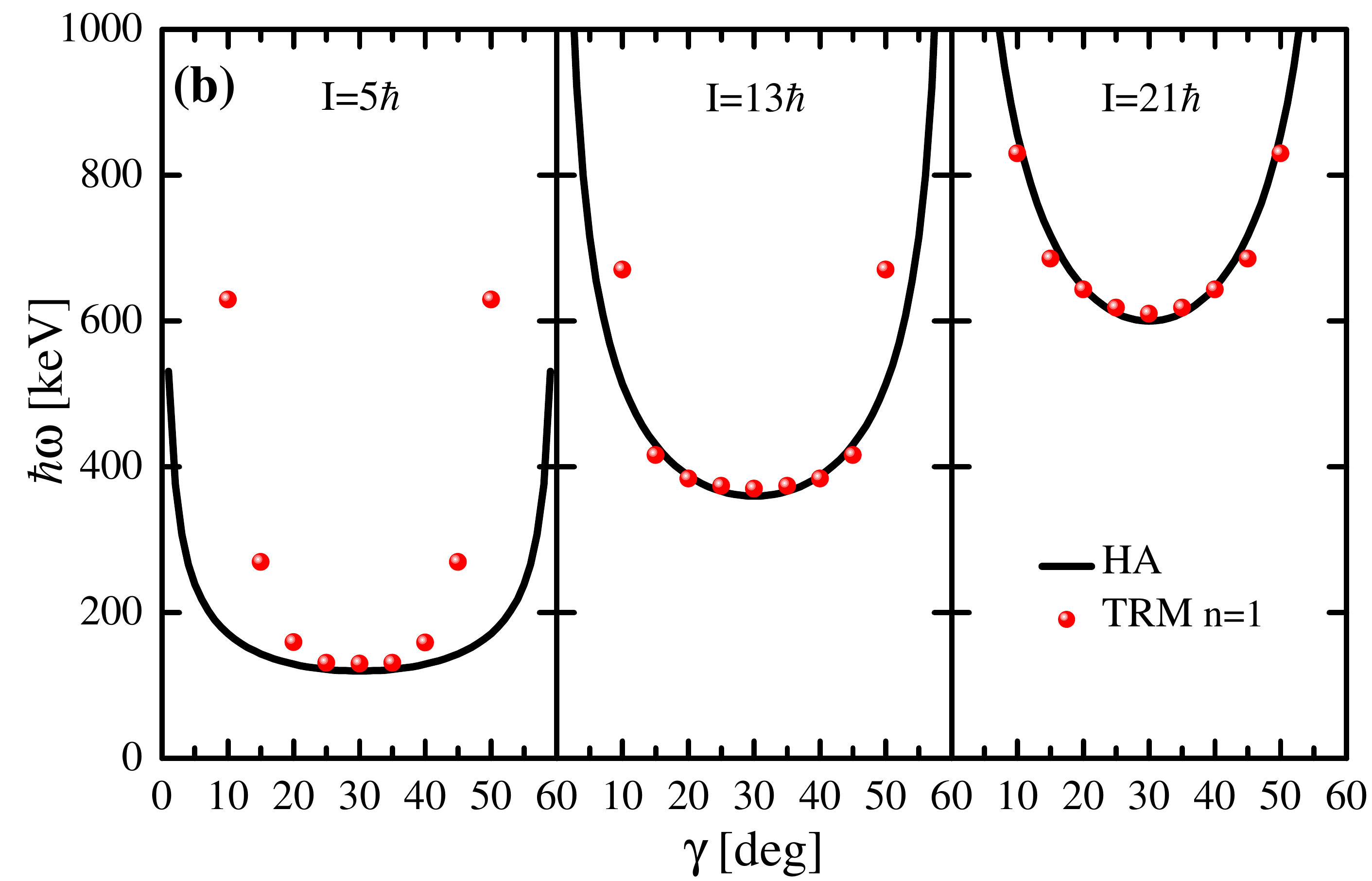}
\caption{(a) The hydrodynamical  MoI of the three principal axes (denoted by
$k =1,2,3$) as functions of $\gamma$. The unit is taken as ${\cal J}_{0}$.
(b) The wobbling frequency as functions of the $\gamma$ calculated by
HA formula Eq.~(\ref{eqiHAwob}) and TRM using Eq.~(\ref{eq2}) with $n=1$.
\label{figMOI} }
\end{figure}

The moment of interia (MoI) is a key parameter to describe the
wobbling excitation. The hydrodynamical MoI is very reasonable for
the triaxial deformed nuclei, and is consistent with
cranking shell model~\cite{Frauendorf182}. In Fig.~\ref{figMOI}(a),
we present the hydrodynamical MoI of the three principal
axes~\cite{Ring},
\begin{equation}\label{eqMOI}
 {\cal J}_{k}= {\cal J}_0 \sin ^{2}(\gamma-\frac{2}{3} \pi k),
\end{equation}
with $\gamma$ the triaxial deformation parameter and the unit of
${\cal J}_{0}$. In the range of $0\leq\gamma\leq\pi/3$, $k=1,2,3$ corresponds to
the intermediate ($m$-), short ($s$-) and  long ($l$-) axis, respectively. Obviously,
${\cal J}_1$, i.e., the MoI of the $m$-axis, is the largest.

The harmonic approximation (HA) for the wobbling excitation and the theoretical framework of  triaxial rotor model~(TRM) can be found in Ref.~\cite{Bohr75}. Using the hydrodynamical MoI, the wobbling frequency calculated by HA formula
\beq \label{eqiHAwob}
\hbar \omega=
I\left[\left(\frac{1}{{\cal J}_{2}}-\frac{1}{{\cal J}_{1}}\right)
\left(\frac{1}{{\cal J}_{3}}-\frac{1}{{\cal J}_{1}}\right)\right]^{1
/ 2},
\eeq
as functions of the $\gamma$ for $I=5,~ 13,~21\hbar$ are shown in Fig.~\ref{figMOI}(b). Here, we take a value of ${\cal J}_0=100~\hbar^2$MeV$^{-1}$,
which is slightly larger than $\sim70~\hbar^2$MeV$^{-1}$ in
$^{163}$Lu~\cite{Frauendorf14}.  The wobbling frequency is the
smallest at $\gamma=30^{\circ}$, and increases as the $\gamma$
deviates from $30^{\circ}$. It increases dramatically
for $\gamma<10^{\circ}$ or $\gamma>50^{\circ}$. The $\hbar \omega$ value
is direct proportion to ${1}/{\cal J}_0$. Thus, if
${\cal J}_0$ takes value of $20~\hbar^2$MeV$^{-1}$ (suitable for $^{135}$Pr~\cite{Frauendorf14}),  $\hbar \omega$ will be five times as large as
these values in Fig.~\ref{figMOI}(b).

For comparison, the $\hbar\omega$ extracted from TRM are also shown, and
the $\hbar\omega$ in HA  becomes better in agreement with the frequency extracted from TRM
when $\gamma$ is closer to $30^{\circ}$ and spin is larger.

As shown in Fig.~\ref{figMOI}(b), the $\gamma$ degree of freedom is very
important in determining the properties of triaxial nuclei. To examine
the quality of HA, we calculate the results of all $\gamma$ values
systematically in the TRM. As is known, the nucleus described as
$(\beta_2, \gamma)$ have the identical shape with $(\beta_2, -\gamma)$,
$(\beta_2, \pm \gamma  \pm120^\circ)$, where $\beta_2$ is the
quadruple deformation. Thus only the results in the $\gamma$ ranging
from $0^\circ$ to $60^\circ$ are sufficient for discussion. Moreover,
due to the symmetry of ${\cal J}_{k}$ with respect to $\gamma=30^{\circ}$
as shown in Fig.~\ref{figMOI}, the results for $30^{\circ}+\Delta\gamma$
(55$^{\circ}$ to 35$^{\circ}$) are identical to the corresponding
results for $30^{\circ}-\Delta\gamma$ $(5^\circ$ to $25^\circ)$. Thus
we can only focus on the results for $\gamma$ from $0^{\circ}$ to
30$^{\circ}$.

\begin{figure*}[ht!]
\includegraphics[width=13 cm]{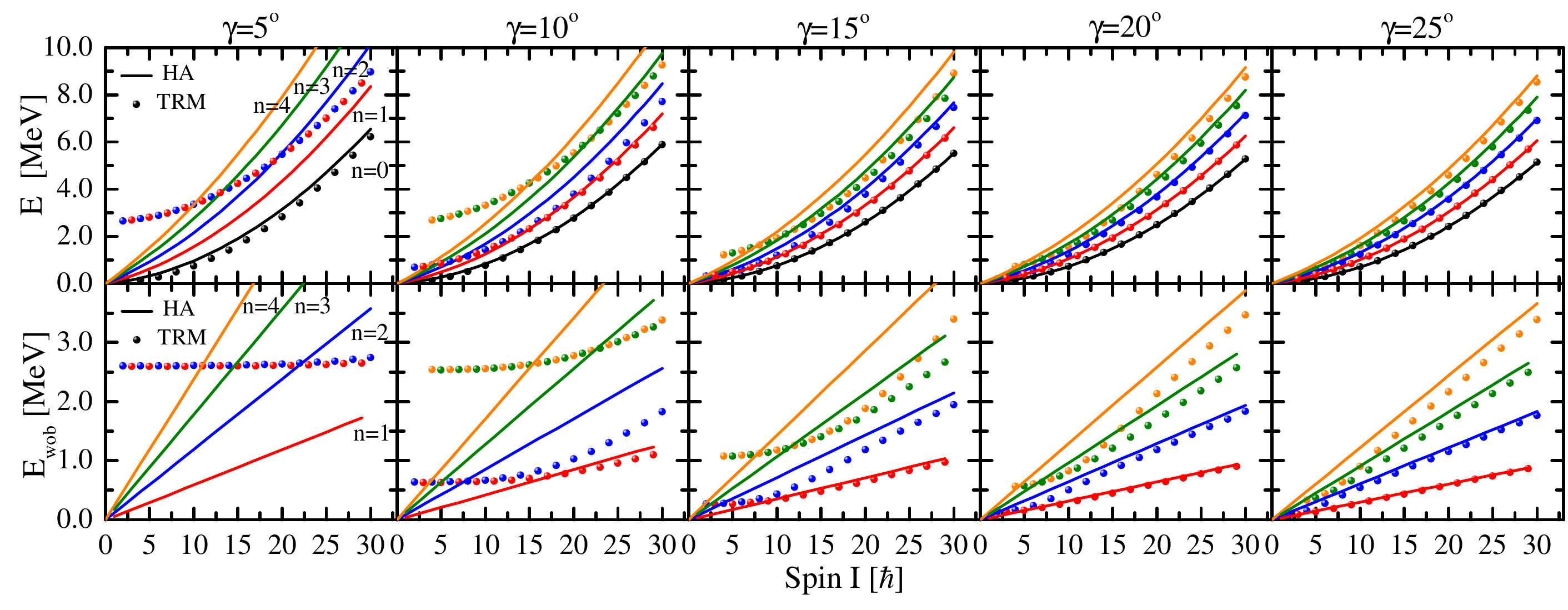}
\includegraphics[width=13 cm]{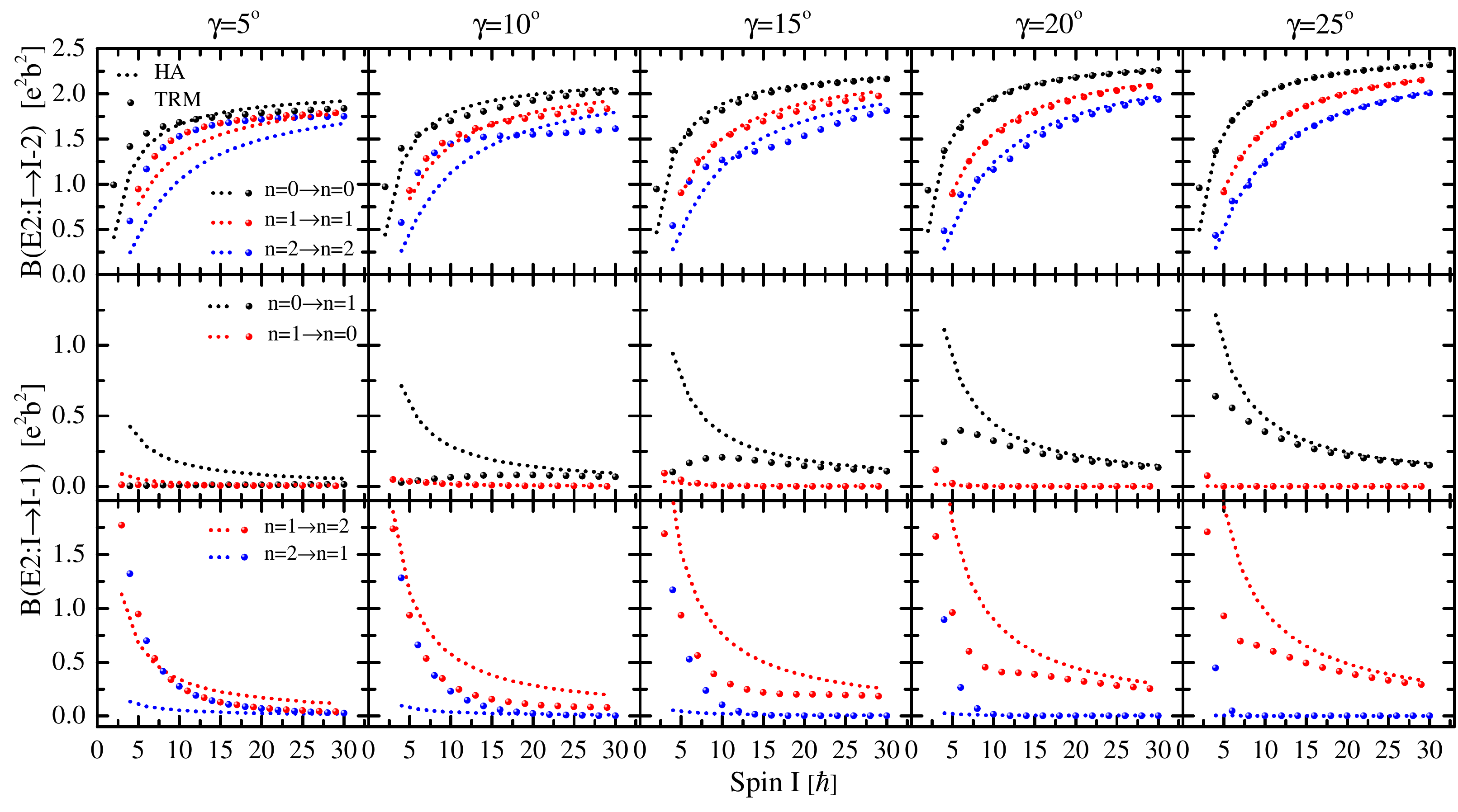}
\caption{Upper panels: The energy spectra and wobbling energies for several lowest
bands calculated by TRM (dots) compare with those by HA formulas (lines)
for $\gamma$ changing from $5^{\circ}$ to $25^{\circ}$.
Lower panels: The intraband and interband $B(E2)$ values for $n=0$, 1,
and $2$ bands calculated by TRM compare with those by HA formulas.
\label{fig2}}
\end{figure*}

\begin{figure*}[ht!]
\subfigure{\includegraphics[width=7.5 cm]{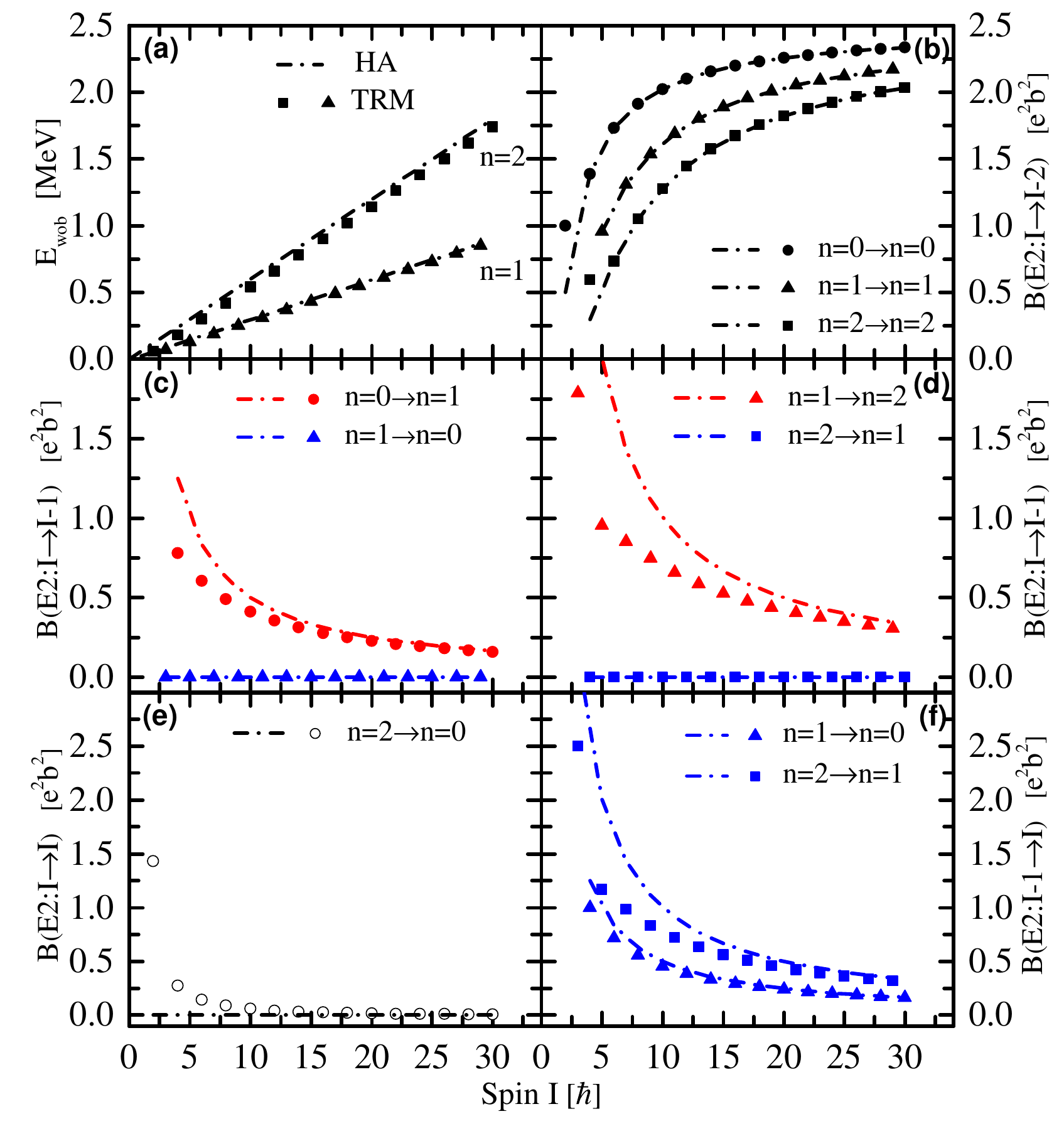}}~~
\subfigure{\includegraphics[width=7.7 cm,bb= 0 -15 310 310]{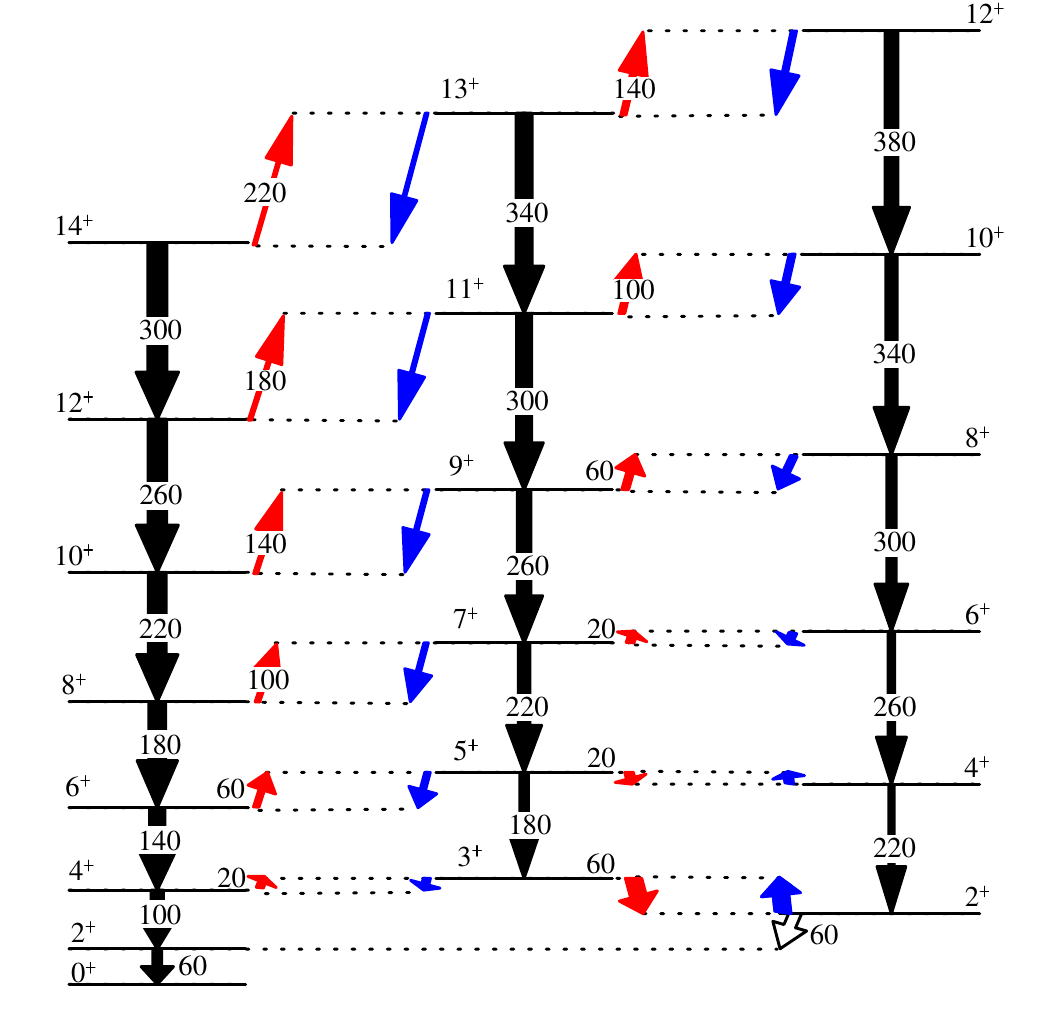}}
\caption{\label{figgam30}
Left panels: The wobbling energies, the intraband and interband $B(E2)$
values for the ground band and  $n=1, 2$ wobbling bands calculated by TRM with
$\gamma=30^{\circ}$ compare with those in HA.
Right panels: The energy level scheme calculated by the TRM for the ground
band and $n=1, 2$ wobbling bands. The
transition energy are denoted, and the thickness of the transitions
is proportional to $B(E2)$ values.}
\end{figure*}

The energy spectra, wobbling energies, as well as the reduced electric
quadrupole transition probabilities as functions of spin for the several
lowest bands calculated by TRM for $\gamma$ changing from $5^{\circ}$ to
$25^{\circ}$ are shown in Fig.~\ref{fig2} and for $\gamma=30^\circ$ in
Fig.~\ref{figgam30}, in comparison with those obtained by the HA
formulas. The energy spectra are obtained by diagonalizing
the TRM Hamiltonian~\cite{Bohr75},
\beq\label{Hamitonian}
 \hat{H} = \sum_{k=1}^{3} \frac{\hat{I}_{k}^{2}}{2 \mathcal{J}_{k}}
 =  A_{1}\hat{I}_{1}^{2}+A_{2}\hat{I}_{2}^{2}+A_{3}\hat{I}_{3}^{2},
\eeq
with $A_k=1/(2\mathcal{J}_k)$. The wobbling energies, defined as the
energy differences between the excited states and the yrast state, are
extracted as~\cite{CQB15}
\begin{eqnarray}\label{eq1}
 E_{\textrm{wob}} =E(n, I)-E(0, I),
\end{eqnarray}
for even $n$ bands, and
\begin{eqnarray}\label{eq2}
 E_{\textrm{wob}} = E(n, I)-\frac{1}{2}[E(0, I-1)+E(0, I+1)],
\end{eqnarray}
for odd $n$ bands, in which
$E(n, I)$ denotes the energy of spin $I$ in the $n$-th excited band. The reduced electromagnetic
transition probabilities are calculated by the operator~\cite{Bohr75}
\begin{align}
{\cal \hat{M}}(E2, \mu) =\sqrt{\frac{5}{16\pi}}\hat{Q}_{2\mu},
\end{align}
with the obtained eigen TRM wave functions. The quadrupole moments
in the laboratory frame ($\hat{Q}_{2\mu}$) and the intrinsic
system ($\hat{Q}'_{2\mu}$) are connected by the relation
\begin{align}
\hat{Q}_{2\mu}&=\mathcal{D}_{\mu 0}^{2 *} \hat{Q}_{20}^{\prime}
+\left(\mathcal{D}_{\mu 2}^{2 *}+\mathcal{D}_{\mu -2}^{2 *}\right) \hat{Q}_{22}^{\prime}\nonumber\\
&=\mathcal{D}_{\mu 0}^{2 *} Q\cos \gamma
+\left(\mathcal{D}_{\mu 2}^{2 *}+\mathcal{D}_{\mu -2}^{2 *}\right) \frac{1}{\sqrt{2}}Q\sin \gamma.
\end{align}

For small triaxial deformation $\gamma=5^\circ$ and $10^\circ$,
there is rather large difference between the HA and
TRM results. For $\gamma=15^{\circ}$, the wobbling
energies for $I>10\hbar$ of $n=1$ band in HA are in agreement with those
in TRM, while for $n>1$ bands in HA have large deviation
from TRM results. As the $\gamma$ increasing, the quality of agreement
between TRM and HA becomes better. HA results for both energy spectra
and wobbling energies are in nice agreement with TRM over the
whole spin range for small $n$=1, 2  phonon wobbling bands for
$\gamma=25^{\circ}$. When $\gamma=30^{\circ}$, the HA formulas could
give very good descriptions for the TRM results, which implies that
the rotational axis exhibit a very good harmonic oscillations with
respect to $m$-axis with the largest MoI.

The intraband and interband $B(E2)$ in HA formulas are calculated as~\cite{Bohr75}:
\begin{eqnarray}
&&\label{eqwobBE2}B(E 2; {n} I \rightarrow {n}, I \pm 2) \approx \frac{5}{16 \pi} e^{2} Q_{2}^{2}\\
&&\label{eqwobBE22}B(E 2 ; {n} I \rightarrow {n}-1, I-1)\nonumber\\
&&\quad =\frac{5}{16 \pi} e^{2} \frac{{n}}{I}\left(\sqrt{3} Q_{0} x-\sqrt{2} Q_{2} y\right)^{2}\\
&&\label{eqwobBE23}B(E 2 ; {n} I \rightarrow {n}+1, I-1)\nonumber\\
&&\quad =\frac{5}{16 \pi} e^{2} \frac{{n}+1}{I}\left(\sqrt{3} Q_{0} y-\sqrt{2} Q_{2} x\right)^{2},
\end{eqnarray}
where $Q_0$ and $Q_2$ are the quadrupole moments with respect to the
$m$-axis, and $x =\sqrt{\left[\alpha/(\hbar \omega)+1\right]/2}$,
$y =\sqrt{\left[\alpha/(\hbar \omega)-1\right]/2}$ with
$\alpha \equiv\left(A_{2}+A_{3}-2 A_{1}\right) I$.
The quadrupole moment $Q=\sqrt{e^2Q_0^2+ e^2Q_2^2}$ takes values of
$\sqrt{16\pi}$ $e$b in the calculations, which is close to the value
of $\sim$9 $e$b in Lu isotopes~\cite{Lu1631}.

For the intraband $B(E2,I \rightarrow I-2)$ values, the HA results
given by Eq.~(\ref{eqwobBE2}) are constants, which are independent of
spin $I$ and wobbling phonon number. This equation results from
the approximation of $\langle I,K,2,-2|I-2,K^\prime\rangle\approx1$~\cite{Bohr75}.
Here, we restore the approximation by adding the square of CG coefficient
$\langle I,K,2,-2|I-2,K^\prime\rangle^2$, with $K = I -n$, $K^\prime = I-2 -n$.
The values of $K$ and $K^\prime$ are taken based on the wobbling picture with
$\gamma=30^\circ$. A similar recipe for $n=0$, $1$ bands was already made in
Ref.~\cite{Hagemann05}. After such modifications, the HA formula could describe well
the characteristics of TRM results, which show the increasing trend of
intraband $B(E2)$ as the increase of wobbling phonon number.

For interband $B(E2,I \rightarrow I-1)$ values, the HA results exhibit
a decreasing trend with respect to spin,
which are determined by the factor $1/I$ in the HA formulas
Eqs.~(\ref{eqwobBE22}) and (\ref{eqwobBE23}). The
$B(E2;n,I \rightarrow n-1,I-1)$ are very small over
almost the whole spin region. For each $\gamma$ under our discussion,
the strength of the interband $B(E2;n,I \rightarrow n+1,I-1)$ is
smaller than that of the intraband $B(E2;n,I \rightarrow n,I-2)$ in
the high spin region by a factor of order $n/I$~\cite{Bohr75}.
Again, the agreement between HA values and TRM results becomes
better as the increase of $\gamma$.

From both the comparisons of the energy spectra and electric quadrupole
transition probabilities of the HA and TRM, it is found that the
agreement are very nice for $\gamma$ changing from $\sim 25^{\circ}$
to  $\sim 35^{\circ}$ over the whole spin range for $n$=0, 1, and 2
bands.

\subsection{$K_m$ of wobbling motion}

The above discussion to judge the quality of HA is based on the observable
of energy and electric quadrupole transition probability. We further
analyze the information of angular momentum to understand this question.
For this purpose, the root mean square of projection of total angular
momentum along the $m$-axis, namely $\langle K_m^2
\rangle^{1/2}$ are calculated in TRM as
\begin{align}
  \langle K_m^2 \rangle^{1/2}
  &= \langle I M |\hat{I_{1}}^{2}|I M \rangle^{1/2}\notag\\
  &=\langle I M |(\hat{I}_++\hat{I}_-)^2/4|I M \rangle^{1/2}.
\end{align}
Here, the $|I M \rangle$ is the eigen wave function of TRM,
\beq\label{eqTRMwave}
 |I M \rangle=\sum_{K \geq 0} C_{I K}|I M K+\rangle,
\eeq
expanded on the basis
\begin{equation}\label{base}
|I M K+\rangle=\sqrt{\frac{2 I+1}{16 \pi^{2}(1+\delta_{K 0})}}
\left[ \mathcal{D}_{M K}^{I}+(-1)^{I}\mathcal{D}_{M -K}^{I}\right].
\end{equation}

\begin{figure}[ht!]
~\includegraphics[width=8.05cm]{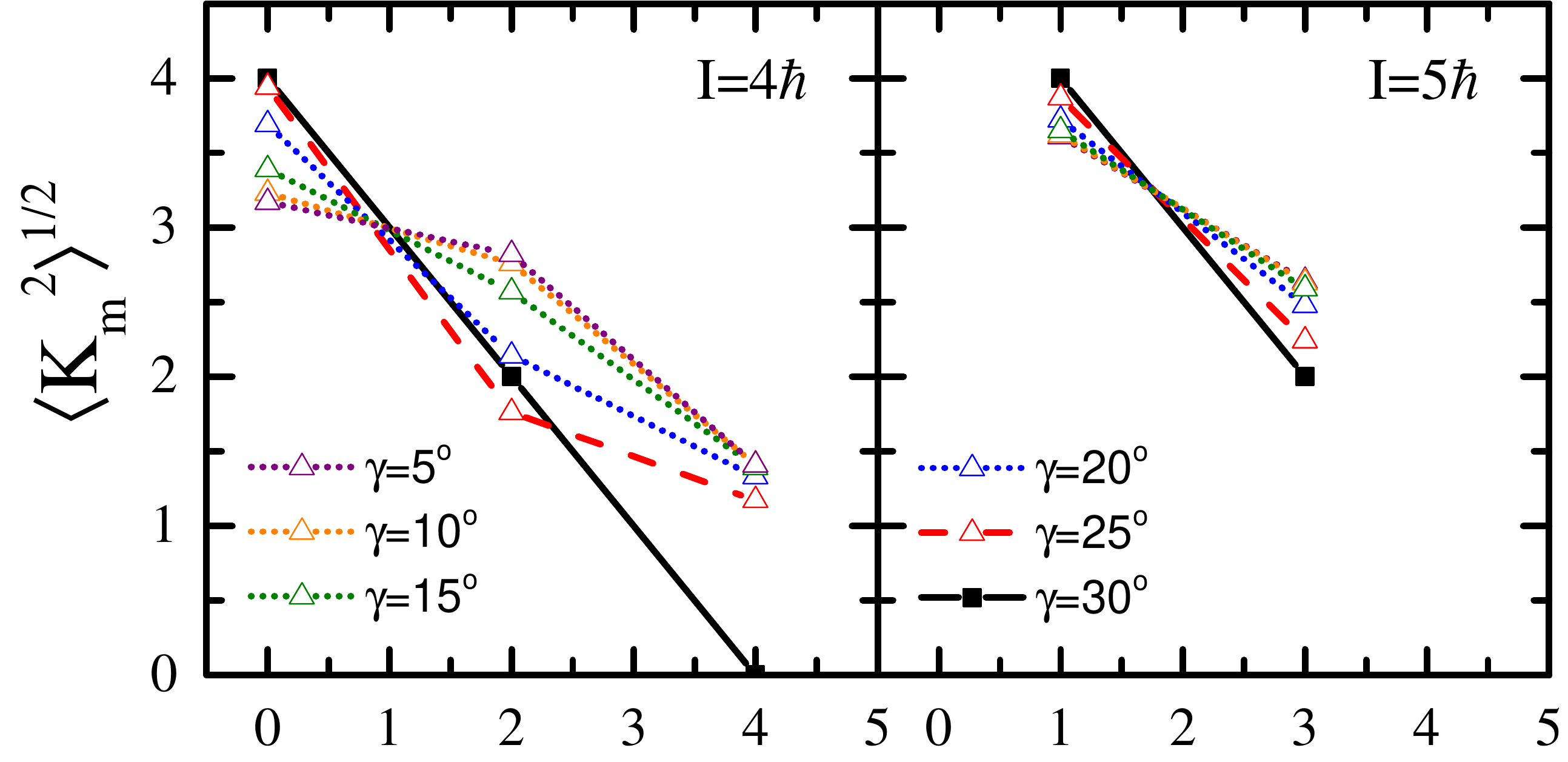}\\
\includegraphics[width=8.1cm]{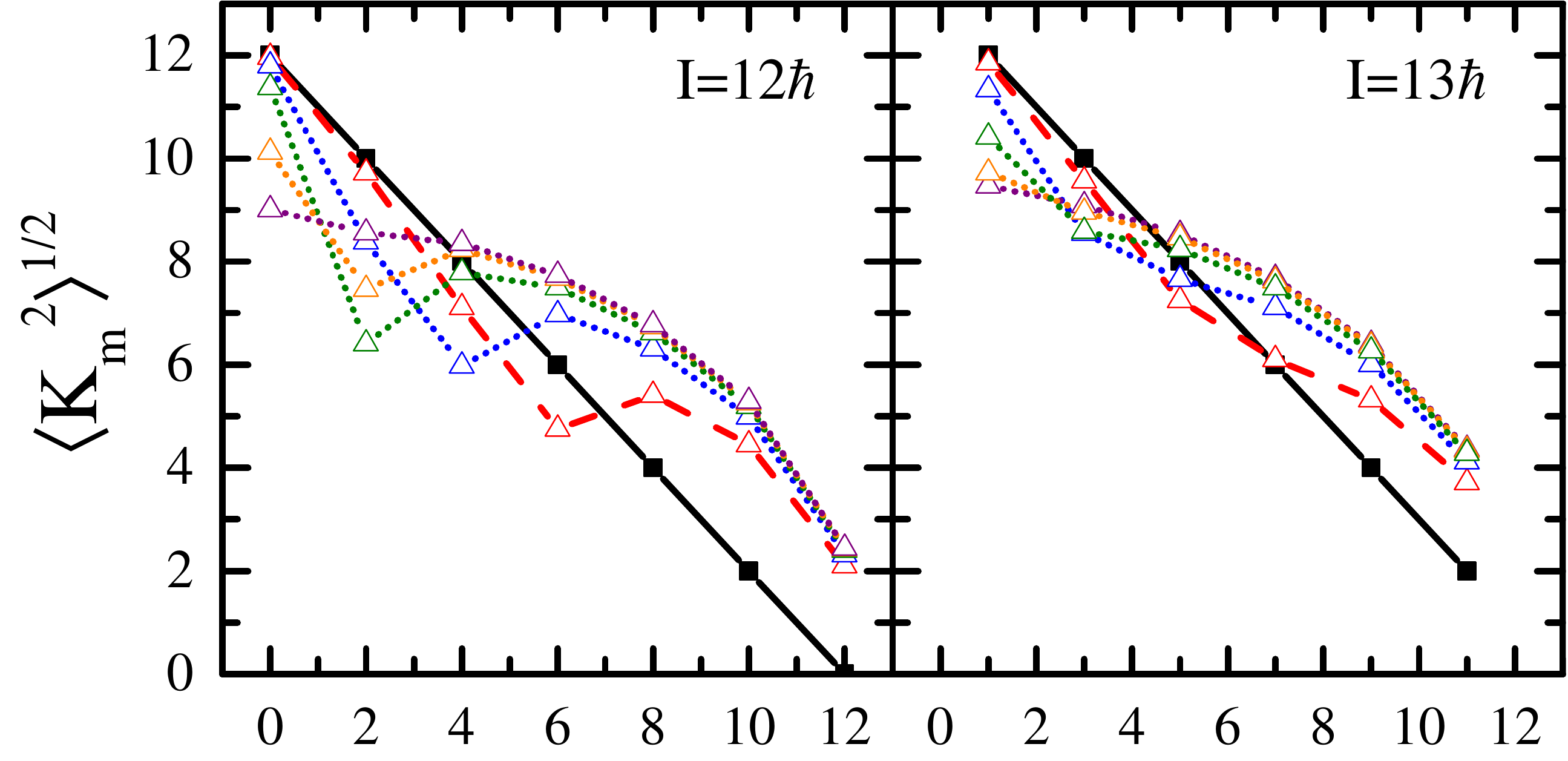}\\
\includegraphics[width=8.1cm]{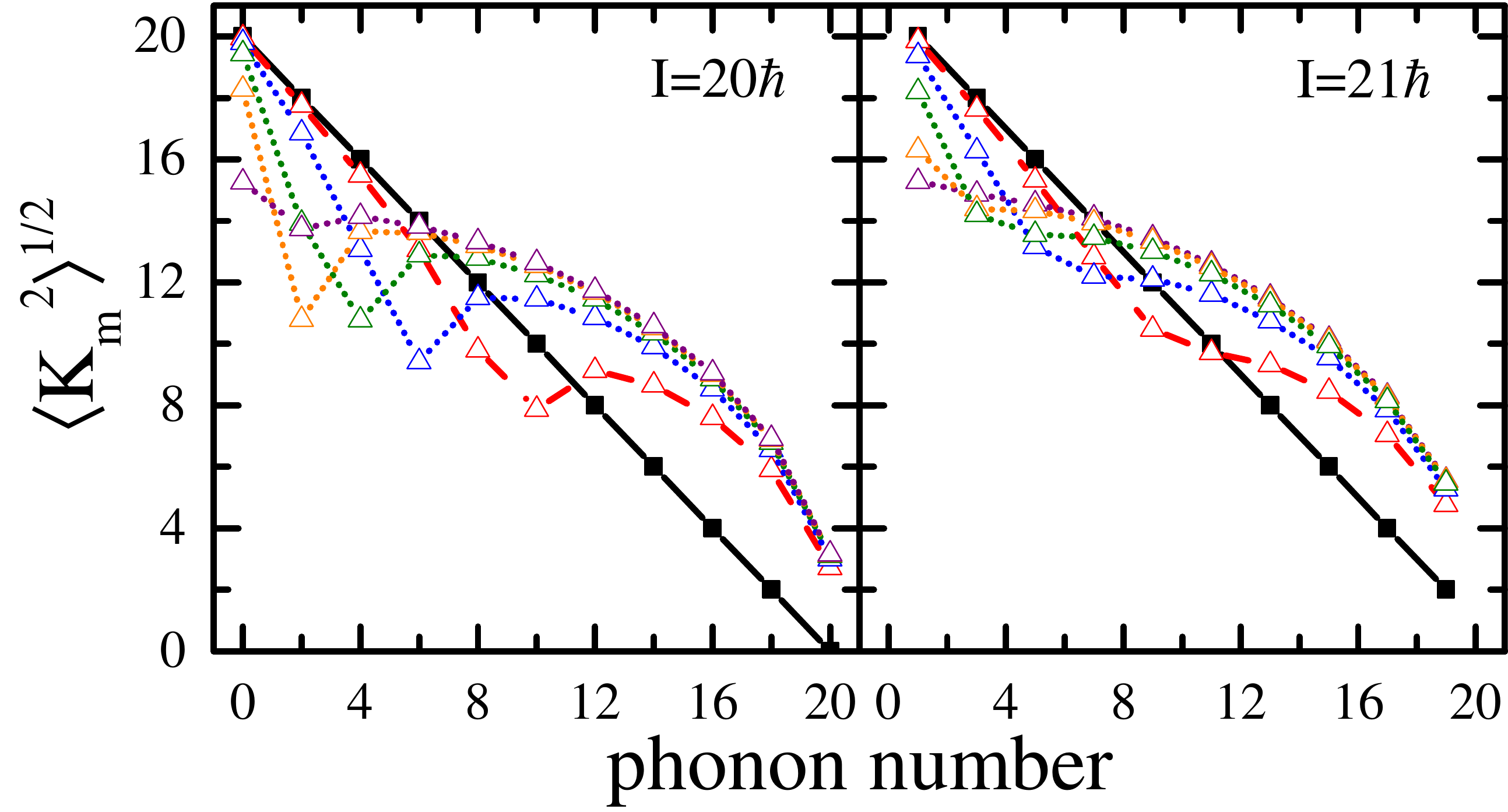}
\caption{The root mean square of the projections of total angular momentum
on the $m$-axis ($K_m$) with $\gamma$ changing from $5^{\circ}$
to $30^{\circ}$ for $I=4, 5, 12, 13, 20, 21\hbar$.\label{figK} }
\end{figure}

The root mean square of  $K_m$ with $\gamma$ changing from
$5^{\circ}$ to $30^{\circ}$ for  $I=4$, 5, 12, 13, 20, $21\hbar$
are shown in Fig.~\ref{figK}. It is found that the relationship
$\langle K_m^2 \rangle^{1/2} =I-n$ with $n=1$, ... , $n_{\textrm{max}}$
are satisfied strictly for $\gamma=30^{\circ}$. The differences between
$\langle K_m^2 \rangle^{1/2}$ and $I-n$ will increase if $\gamma$
gradually deviates from $30^{\circ}$, or if the phonon number
$n$ gradually increases.

The above variation as  $\gamma$ and $n$ can be understood by the
$K_m$ structure, namely the probability distribution of different
$K_m$ values. As examples, the $K_m$ structure for all states with different phonon
number at spin $12\hbar$ and $13\hbar$ are shown in Table~\ref{Tab:I12}.

Let us first investigate the case of $\gamma=30^\circ$. It is
much easier to express the results  from $\gamma=90^\circ$, which
has the identical shape with $\gamma=30^\circ$ except the
$m$-axis is chosen as the quantum 3-axis.
Due to ${\cal J}_1={\cal J}_2$, ${A}_1={A}_2=4A_3$ in
Eq.~(\ref{Hamitonian}) when $\gamma=90^\circ$, the Hamiltonian
reads now
\begin{eqnarray}\label{eqhami90}
 \hat{H} &=& \frac{1}{2}(A_{1}+A_{2})(\hat{\bm{I}}^{2}-\hat{I}_{3}^{2})
   +A_{3}\hat{I}_{3}^{2}.
\end{eqnarray}
Thus the projection $K_m$ is good quantum number. From the calculated
results the following relationship are satisfied strictly for
$n=1$,..., $n_{\textrm{max}}$,
\beq
 K_m =I-n.
\eeq

When $\gamma$ deviates from $\gamma=30^\circ$, the non-diagonal term
in Eq.~(\ref{Hamitonian}) will introduce the $K$-mixing, $K_m$ is not
a good quantum number. As shown in Table~\ref{Tab:I12}, for $n=0$ ground
state of 12$\hbar$, the components of $K_m=12$ is over $90\%$ when
$\gamma=25^{\circ}$ and $20^{\circ}$, and decreases to $76\%$ when
$\gamma=15^{\circ}$. For $n=2$ phonon state, the component of
$K_m=I-2=10$ is the dominant only for $\gamma=25^\circ$.

\begin{table*}
\centering
 \caption{\label{Tab:I12} $K_m-$structure for $I=12$ and $13\hbar$}
\begin{tabular}{c|c|c|c|ccc}
 \hline
 \hline
 $I=12\hbar$  & $\gamma=30^{\circ}$ & $\gamma=25^{\circ}$ & $\gamma=20^{\circ}$ &  $\gamma=15^{\circ}$ \\
\hline
n=0 &  \blue{$100\% |12\rangle  $} &  \blue{$98\% |12\rangle$}+ $2 \% |10\rangle $
& \blue{$90\%|12\rangle$}+$8\%|10\rangle+1\%|8\rangle$  ~~& \blue{$76\% |12\rangle$}+$15\%|10\rangle+5\%|8\rangle$
 \\
n=2 &  \blue{$ 100\%|10\rangle $}&\blue{$81\%|10\rangle$}+$15\%|8\rangle+2\%|12\rangle$
& $35\%|10\rangle+29\%|8\rangle+15\%|6\rangle$ &  $21\% |2\rangle+20\%|4\rangle+19\%|6\rangle$ \\
n=4 &  \blue{$100\% |8\rangle $}&$38\%|8\rangle+31\%|6\rangle+13\%|10\rangle $
&$29\%|2\rangle+26\%|10\rangle+20\%|4\rangle$
&  $33\% |10\rangle+19\%|2\rangle+19\%|8\rangle$ \\
n=6 &  \blue{$100\% |6\rangle $}&$35\%|2\rangle+24\%|8\rangle+20\%|4\rangle $
&$25\%|10\rangle+23\%|6\rangle+21\%|8\rangle$
&  $37\% |10\rangle+28\%|6\rangle+13\%|0\rangle$ \\
n=8 &  \blue{$100\% |4\rangle $}&$37\%|6\rangle+21\%|8\rangle+20\%|0\rangle$
& $39\%|8\rangle+23\%|4\rangle+15\%|0\rangle$
&  $43\% |8\rangle+26\%|4\rangle+14\%|0\rangle$  \\
n=10 &  \blue{$100\% |2\rangle $}&$47\%|4\rangle+29\%|6\rangle+21\%|0\rangle $
&$39\%|6\rangle+33\%|4\rangle+18\%|0\rangle$
&  $41\% |6\rangle+28\%|4\rangle+17\%|0\rangle$ \\
n=12 & \blue{ $100\% |0\rangle $}&$48\%|2\rangle+38\%|0\rangle+12\%|4\rangle $
&$48\%|2\rangle+34\%|0\rangle+16\%|4\rangle$
 &  $47\% |2\rangle+33\%|0\rangle+17\%|4\rangle$  \\
\end{tabular}

\begin{tabular}{c|c|c|c|cc}
 \hline
   $I=13\hbar$ & $\gamma=30^{\circ}$ & $\gamma=25^{\circ}$ & $\gamma=20^{\circ}$ & $\gamma=15^{\circ}$ \\
\hline
n=1 &   \blue{$100\% |12\rangle  $}
&   \blue{$93\% |12\rangle$}+ $7\% |10\rangle $
& \blue{$72\%|12\rangle$}+$21\%|10\rangle+5\%|8\rangle$ &  $47\%|12\rangle+29\%|10\rangle+15\%|8\rangle$\\
n=3 &  \blue{$ 100\%|10\rangle $}
&\blue{$66\%|10\rangle$}+$23\%|8\rangle+7\%|12\rangle$
&$31\%|8\rangle+22\%|6\rangle+20\%|12\rangle$&  $35\%|12\rangle+25\%|6\rangle+17\%|8\rangle$\\
n=5&  \blue{$ 100\%|8\rangle $}
&$36\%|6\rangle+24\%|8\rangle+21\%|10\rangle $
 &$37\%|10\rangle+27\%|4\rangle+16\%|6\rangle$&  $34\%|10\rangle+23\%|4\rangle+15\%|12\rangle$\\
n=7 &  \blue{$ 100\%|6\rangle $}
&$37\%|8\rangle+35\%|4\rangle+19\%|2\rangle $
& $32\%|8\rangle+23\%|10\rangle+22\%|2\rangle$&  $31\%|10\rangle+23\%|8\rangle+22\%|2\rangle$\\
n=9 &  \blue{$100\% |4\rangle $}
&$46\%|6\rangle+36\%|2\rangle+14\%|8\rangle$
& $37\%|6\rangle+31\%|2\rangle+28\%|8\rangle$&  $33\%|8\rangle+31\%|6\rangle+29\%|2\rangle$\\
n=11 & \blue{$100\% |2\rangle $}
&$46\%|4\rangle+42\%|2\rangle+12\%|6\rangle $
&$47\%|4\rangle+33\%|2\rangle+18\%|6\rangle$& $46\%|4\rangle+30\%|2\rangle+21\%|6\rangle$\\
\hline
\hline
\end{tabular}
\end{table*}

Here, we think the probability of the $K_m=I-n$ component larger
than 50\% might be chosen as a reasonable criteria to judge the
quality of HA approximation. Based on this suggested criteria,
wobbling bands are realized perfectly for $\gamma=30^{\circ}$. For
spin $12\hbar$ and $13\hbar$, nice wobbling occurs for $n=1,2,3$
phonon excitation when $\gamma=25^{\circ}$, and for $n=1$ phonon
excitation when $\gamma=20^{\circ}$. As the spin increasing, the HA
wobbling approximation becomes better. For the states of $20\hbar$
and $21\hbar$, the  probability of $K_m=I-n$ larger than 50\% is
$n=1,2,3,4$ phonon excitation when $\gamma=25^{\circ}$. The obtained
conclusion to judge the quality of HA from the probability of the
$K_m=I-n$ component is very consistent with those judgements from
the energy and transition.

\subsection{Level scheme of wobbling band with $\gamma=30^\circ$}

According to the above discussion, the stable large triaxial deformation
is necessary for the realization of the wobbling excitation. For the
realistic even-even nuclei, the stable triaxial deformation is rare in
the ground state~\cite{Moller2006PRL}. A stable rigid triaxial deformation
with $\gamma \in (25^{\circ}, 35^{\circ})$ is indeed a relatively strict condition. 
It might be one reason why the purely collective form
was difficult to be observed in experiment in the past decades.

We would further explore level scheme for $\gamma=30^\circ$, which
shows a very good wobbling picture, with the hypothesis of a stable
rigid triaxial deformation. Bohr and Mottelson discussed the excited
energies of $\gamma=30^\circ$ very briefly in appendix 6B of
textbook~\cite{Bohr75}. In Fig.~\ref{figgam30}, the wobbling energies
and the $B(E2)$ values for the two lowest wobbling bands calculated
by TRM with $\gamma=30^{\circ}$ are shown in comparison with
those from the HA formulas. The HA formulas in panel (e) and
(f) are new deduced in this paper according to the method in the
textbook~\cite{Bohr75} as
\begin{align}
  & B(E 2 ; {n}, I \rightarrow {n}-2, I ) \approx \frac{5}{16 \pi} e^{2} Q_{0}^{2} \\
  & B(E 2 ; {n}, I-1 \rightarrow {n}-1, I ) \notag\\
  &\quad \approx \frac{5}{16 \pi} e^{2} \frac{{n}}{I}
   \left(\sqrt{3}Q_{0}y-\sqrt{2}Q_{2}x\right)^{2}.
\end{align}

In the right panel, we show level scheme from the TRM results of the
ground band and the $n=1$ and $2$ wobbling bands. The values of
transition energies are marked, and the thickness of the transition
is proportional to $B(E2)$ values.

From the level scheme, some interesting relationships are exhibited as
follows,
\begin{enumerate}
  \item $ E(0, I+2)-E(0, I)= E(1, I+5)-E(0, I+6)=E(2, I+8)-E(1, I+9)$ ,
  e.g., 60~keV  is the transition energy
for $(0,2)\rightarrow (0,0)$, $(1,5)\rightarrow (0,6)$
and $(2,8)\rightarrow (1,9)$. Similar for 100~keV and 140~keV.
  \item $E(0, I+2)-E(0, I)=E(1, I-1)-E(1, I-3)=E(2, I-4)-E(2, I-6)$,
  e.g., 220~keV is the transition energy
for $(0,10)\rightarrow (0,8)$, $(1,7)\rightarrow(1,5)$
and $(2,4) \rightarrow (2,2)$. Similar for 260~keV and 300~keV.
  \item $[E(n, I+4)-E(n, I+2)]-[E(n, I+2)-E(n, I)]= 40~\textrm{keV}$.
\end{enumerate}

\begin{figure}[ht!]
~\includegraphics[width=8.cm]{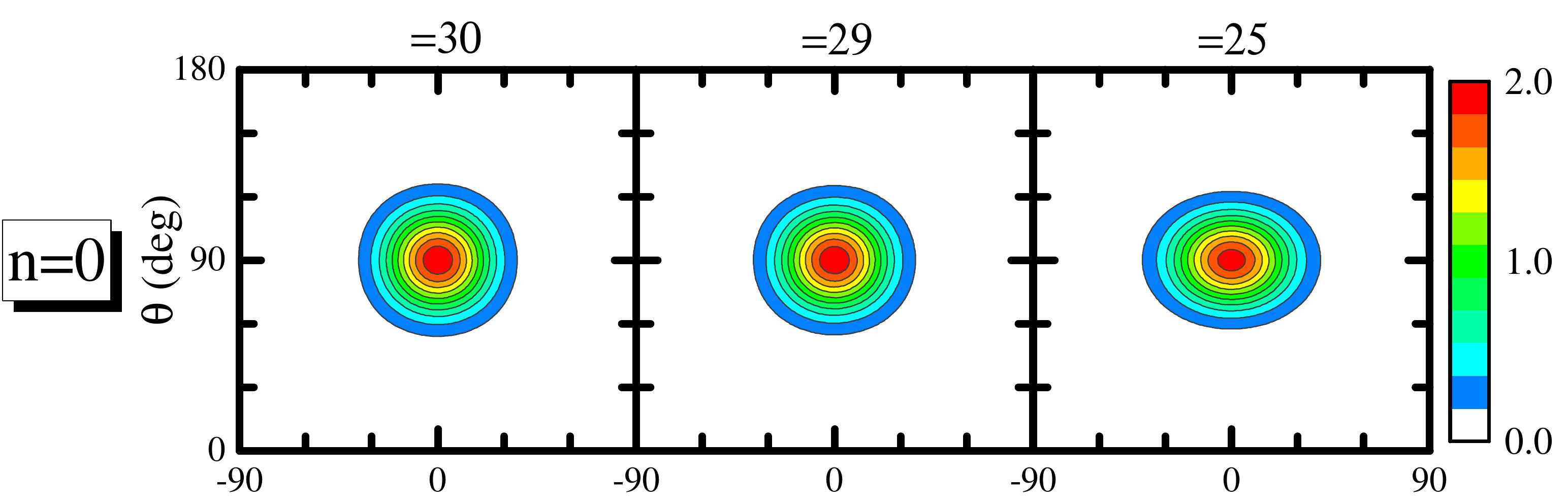}\\
~\includegraphics[width=8.03cm]{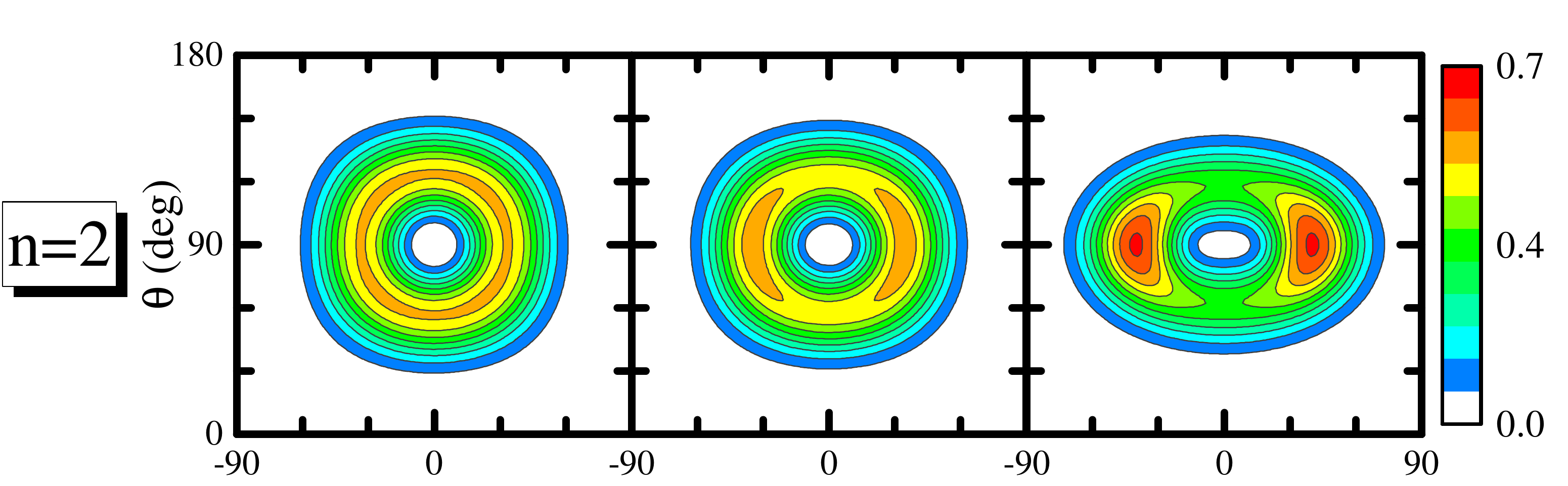}\\
~\includegraphics[width=8.cm]{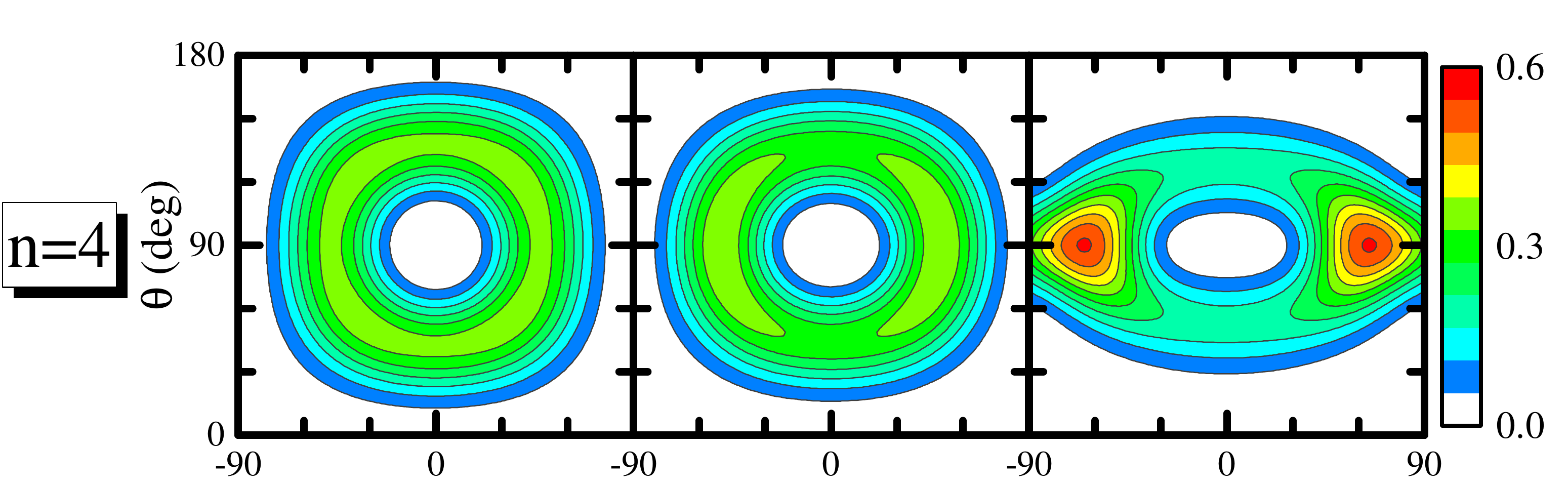}\\
~\includegraphics[width=8.03cm]{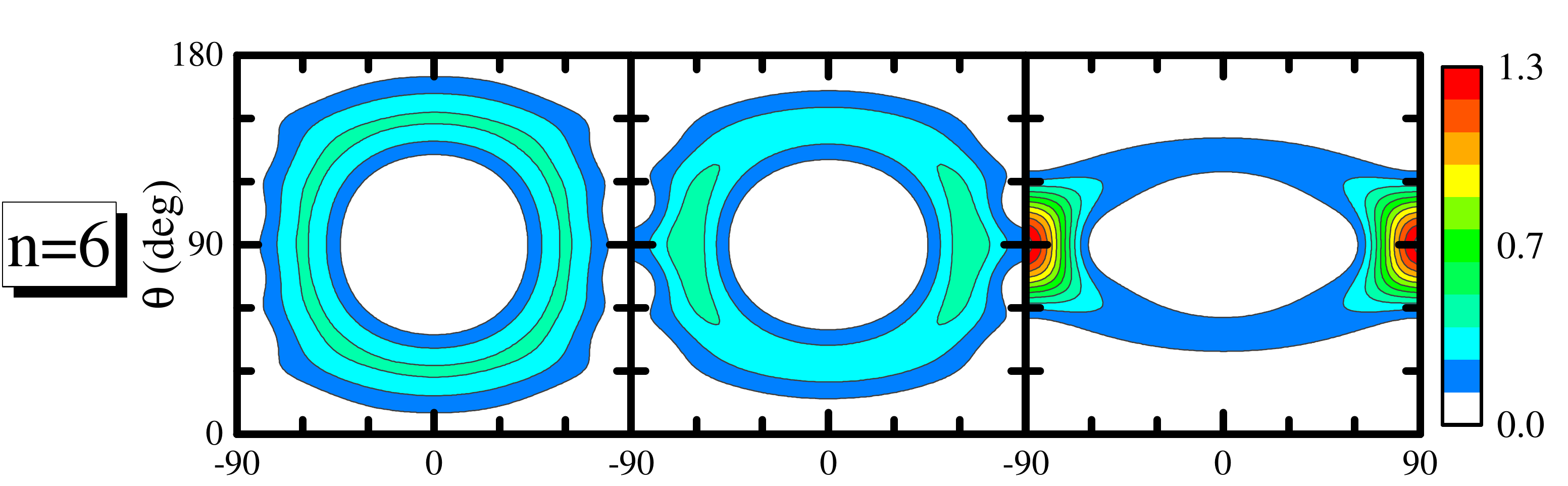}\\
~\includegraphics[width=8.cm]{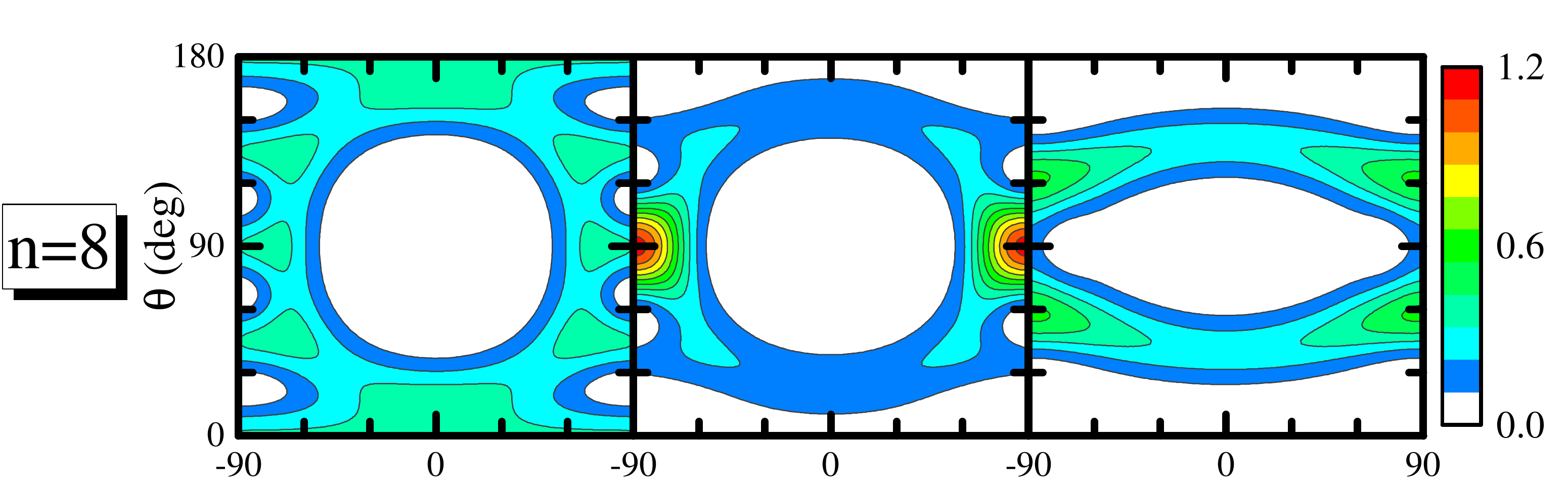}
\includegraphics[width=8.15cm]{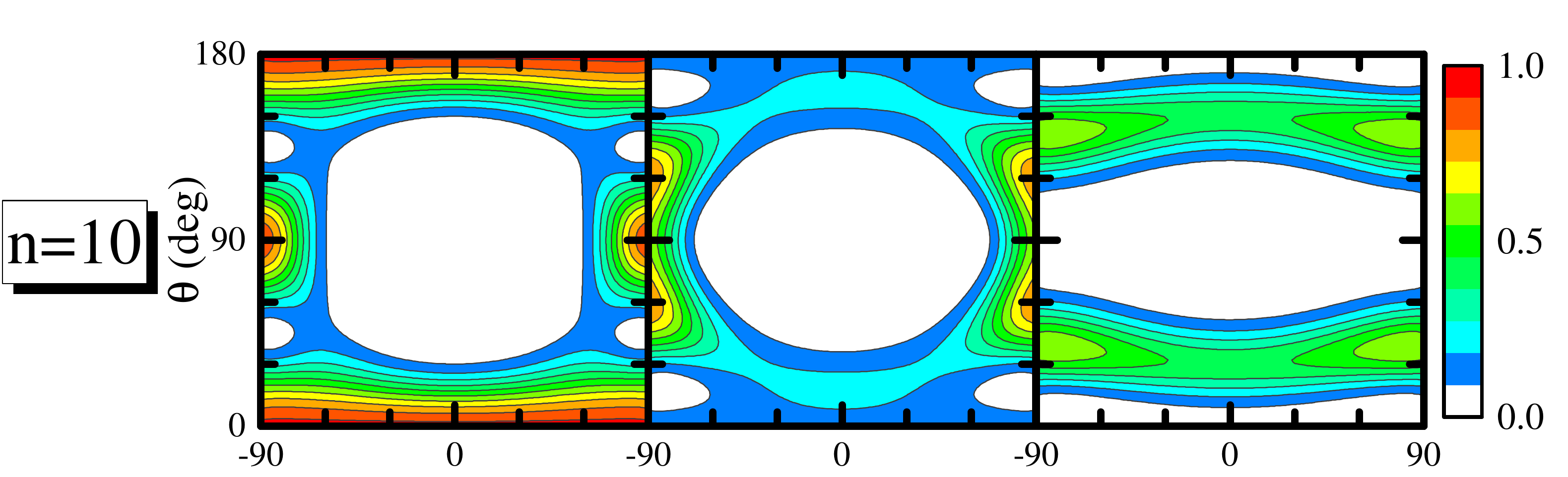}\\
\includegraphics[width=8.15cm]{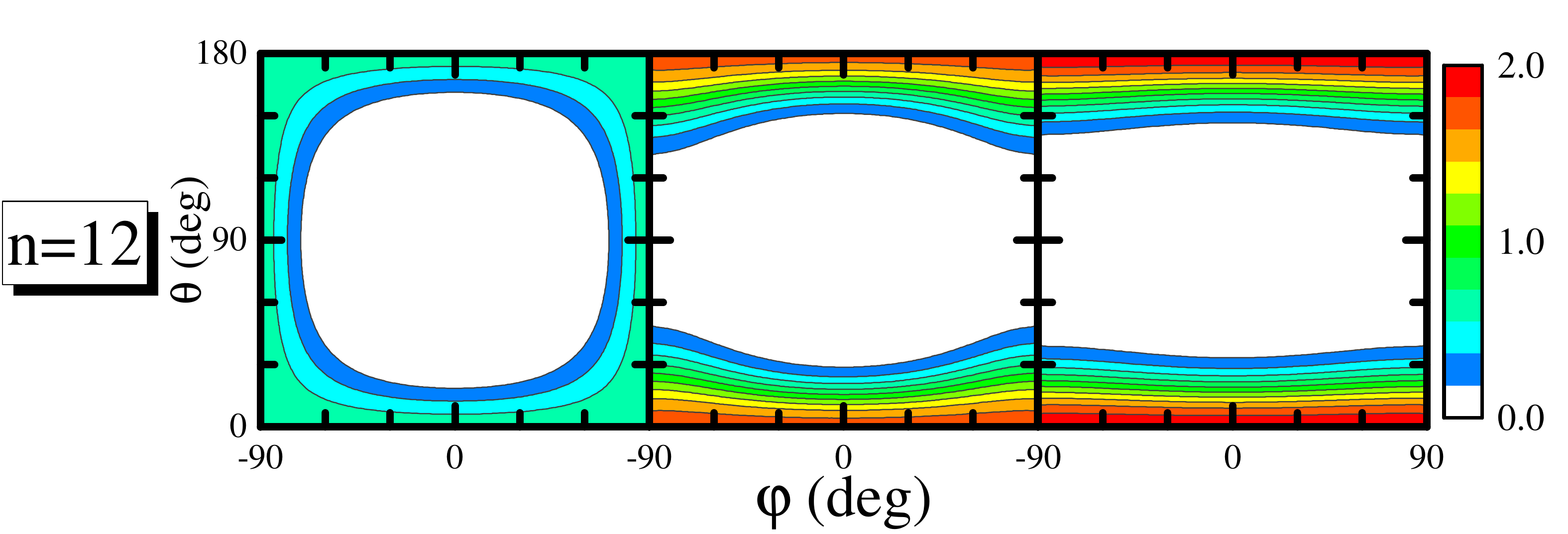}\\
\caption{\label{figAplot} Azimuthal plots for states with phonon
number $n=0$, 2, 4, 6, 8, 10, and 12 at $I=12\hbar$ calculated
by TRM with $\gamma=30^{\circ}$, $29^{\circ}$, and $25^{\circ}$.}
\end{figure}

\begin{figure}[ht!]
\includegraphics[width=4cm]{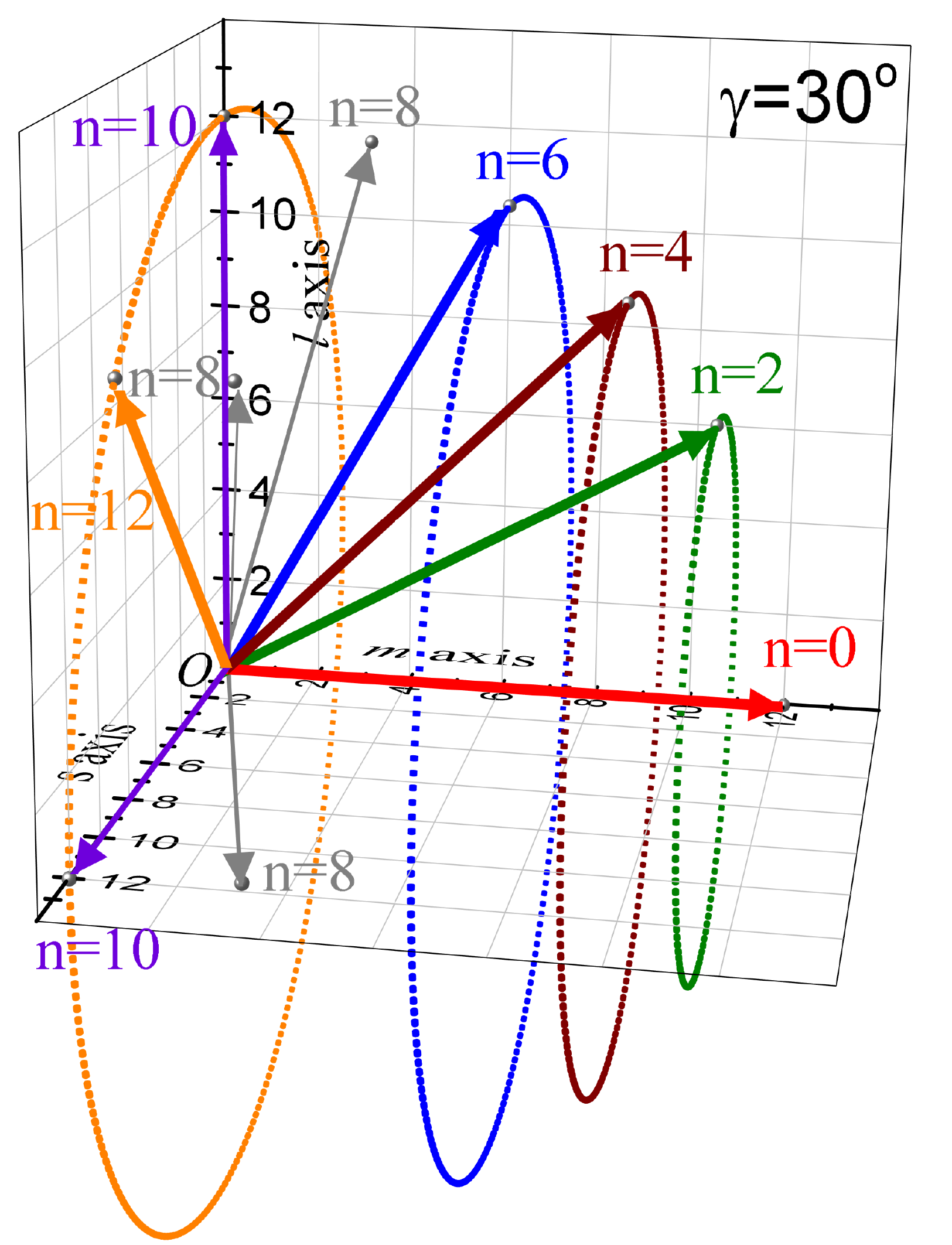}
\includegraphics[width=4cm]{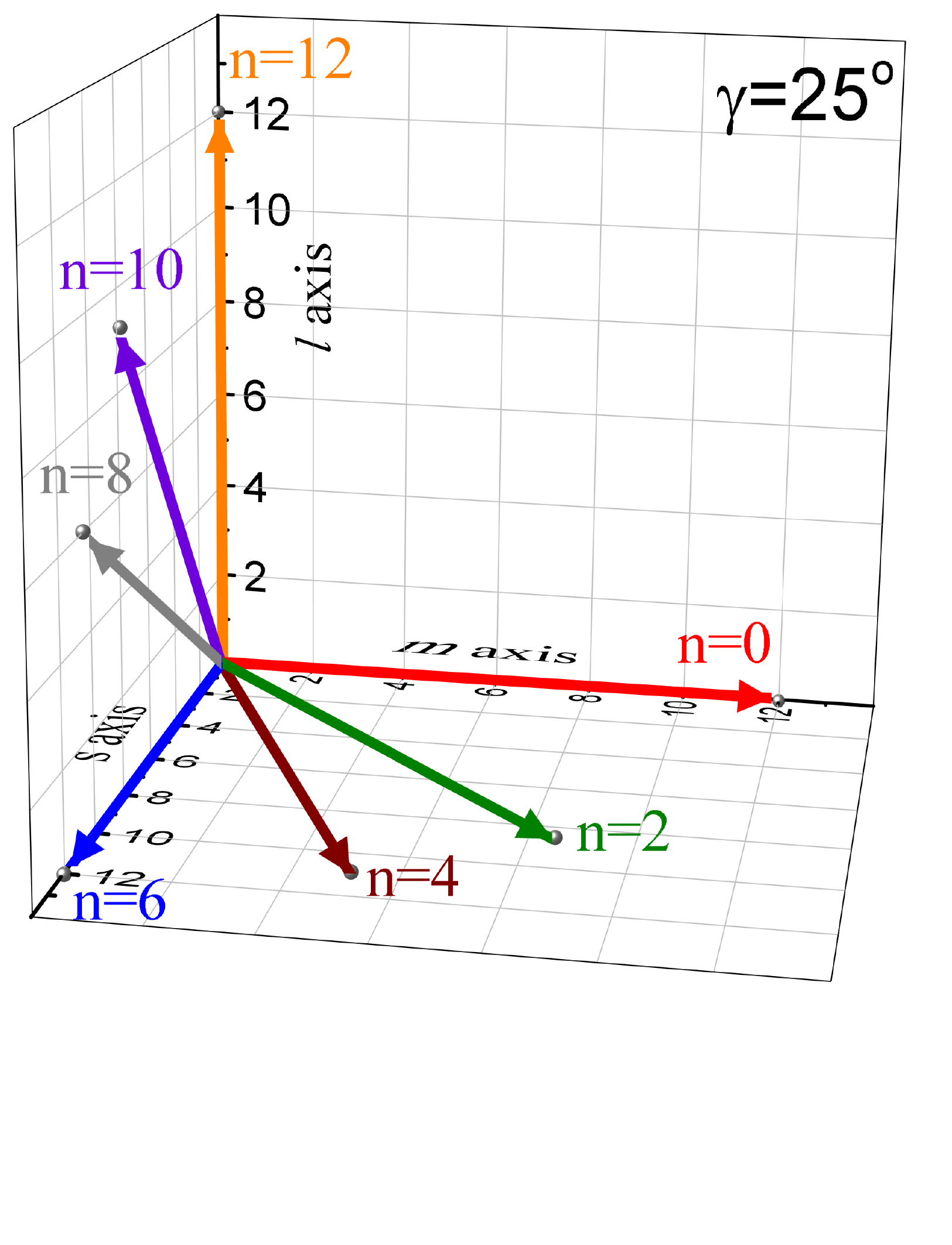}
\caption{\label{figAplot2} Schematic illustration of angular momentum
geometry at spin $I=12\hbar$ for $\gamma=30^{\circ}$ and
$\gamma=25^{\circ}$. The orientation of arrows refer to the maxima in the
azimuthal plots in Fig.~\ref{figAplot}. }
\end{figure}

These relationships are understood as follows. From the Hamiltonian
of TRM in Eq.~(\ref{eqhami90}), one obtains
\begin{eqnarray}\label{eqi3TRM}
 E(n,I)= A_3 I(I+1)+6A_3I(n+\frac{1}{2})-3A_3n^2.
\end{eqnarray}
Alternatively, from the HA formula  $(A_1=A_2=4A_3)$ with
$\hbar\omega=2I\sqrt{(A_{2}-A_{3})(A_{1}-A_{3})}=6I A_3$,
one obtains
\beq\label{eqi3HA}
E(n,I)= A_3I(I+1)+6A_{3}I(n+\frac{1}{2}).
\eeq

Therefore, from either Eq.~(\ref{eqi3TRM}) or Eq.~(\ref{eqi3HA}),
one gets
\begin{eqnarray}
E(n, I+2)-E(n, I)= 4A_3(I+3n)+12A_3,
\end{eqnarray}
and thus the above relationships 2 and 3. Furthermore, from
Eq.~(\ref{eqi3TRM}), one gets
\begin{eqnarray}\label{eqi5TRM}
E(n+1,I+5)-E(n,I+6)= 4A_3(I-3n)+12A_3
\end{eqnarray}
and thus relationship 1. Note that the HA formula Eq.~(\ref{eqi3HA})
can not derive the relationship 1 due to the lack of $-3A_3n^2$ term.
In addition, each values of energy in the level scheme will change according to the rule of $1/{\cal
J}_{0}$  for different  ${\cal J}_0$.

There are one interesting thing worthwhile to be noted from
Fig.~\ref{fig2} and Fig.~\ref{figgam30}(c), (d), the interband
$B(E2,I \rightarrow I-1)$ for  $n=1\rightarrow n=0$ and
$n=2\rightarrow n=1$ are strongly suppressed for both HA and TRM
results. Similar conclusions were obtained in Refs.~\cite{Casten03}.
In the observed wobbling bands in odd-$A$ nuclei, the interband $B(E2,I
\rightarrow I-1)$ exists and links the wobbling excited band and
yrast band, e.g. see Refs.~\cite{Pr135, Lu1632}. It could be
inferred that the linking transitions between the wobbling bands of
even-even nuclei are different from those in odd-$A$ nuclei.

As shown in the level scheme in Fig.~\ref{figgam30}, the $B(E2,I
\rightarrow I-1)$ from $n=0$ to $n=1$ wobbling
band will not occur spontaneously due to it needs
to absorb energy. Such transitions are suggested to be realized by
the method of Coulomb excitation, which has explored the transitions
at the lower spin region of triaxial or octuple deformed nuclei recently~\cite{Nature2013,Doherty17}.

\begin{figure*}[ht!]
\includegraphics[width=8.5cm]{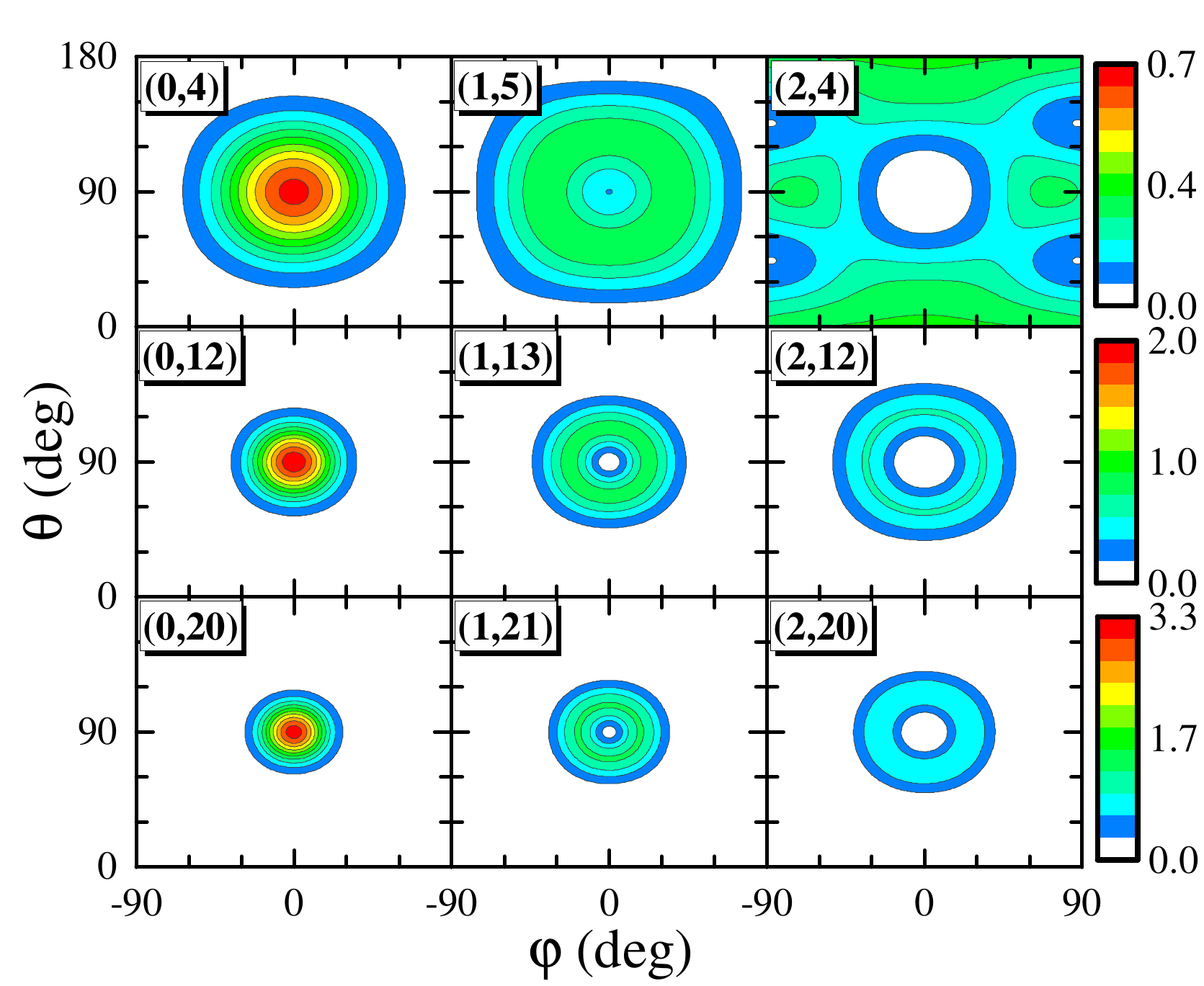}
\includegraphics[width=4.5cm]{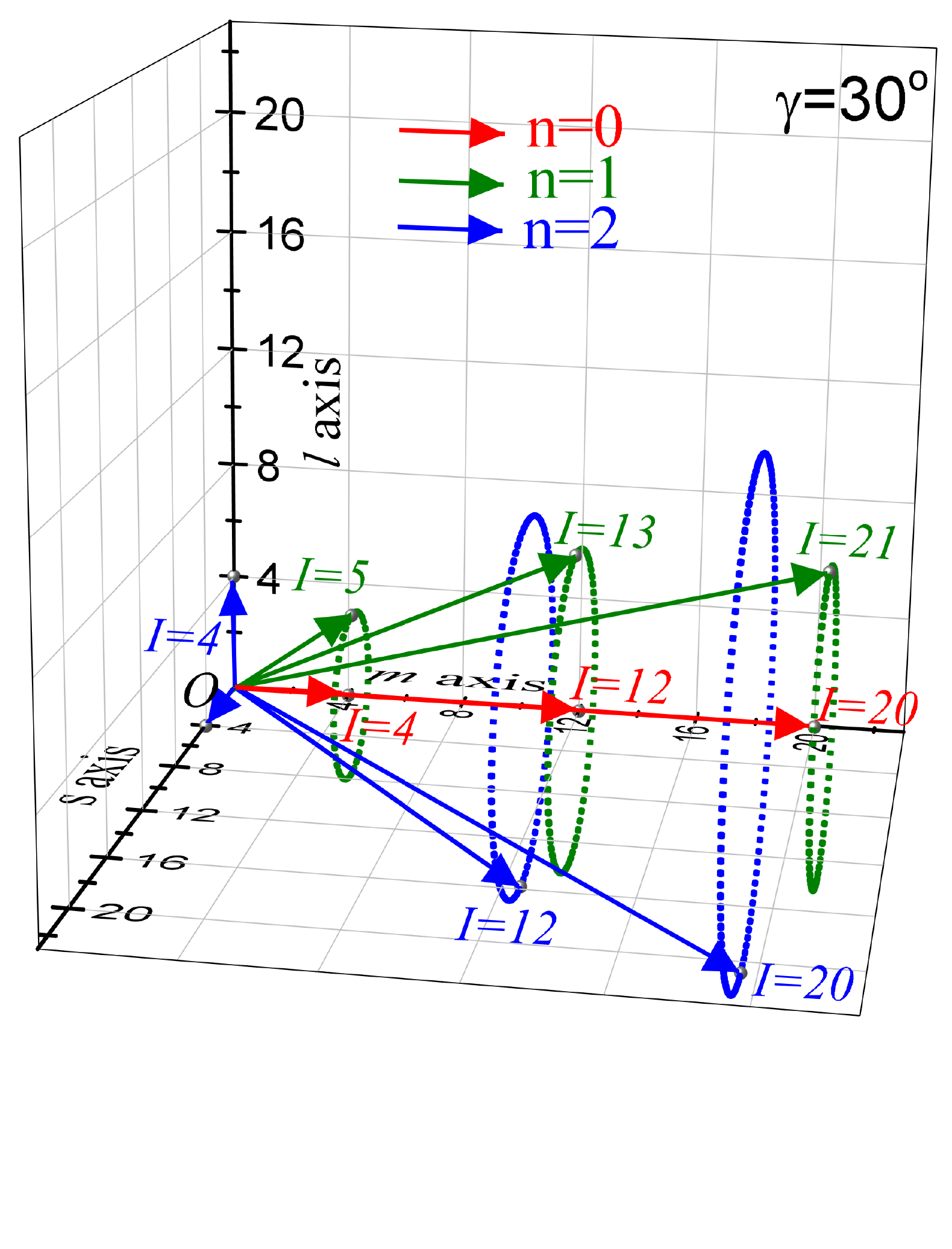}\\
\includegraphics[width=8.5cm]{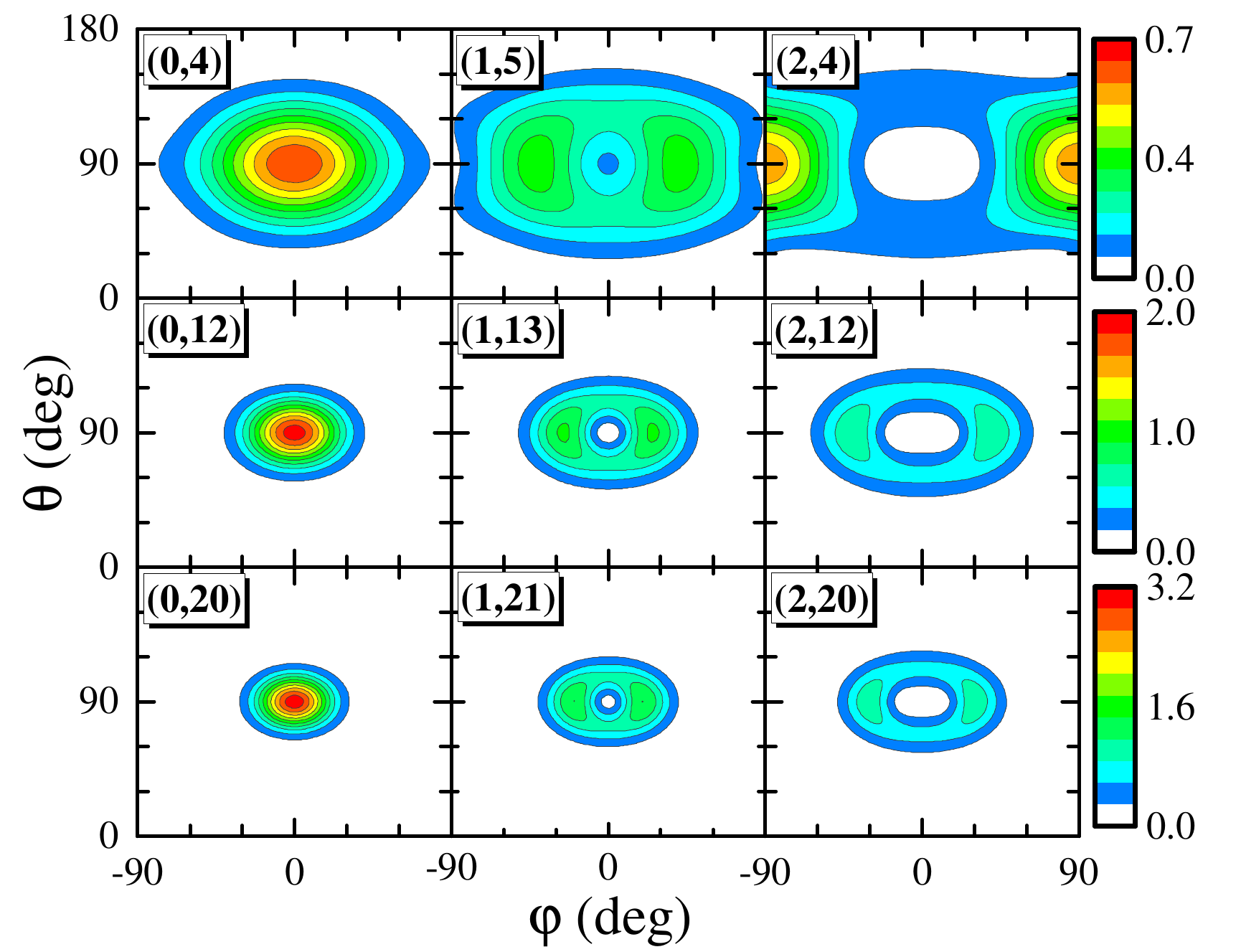}
\includegraphics[width=4.5cm]{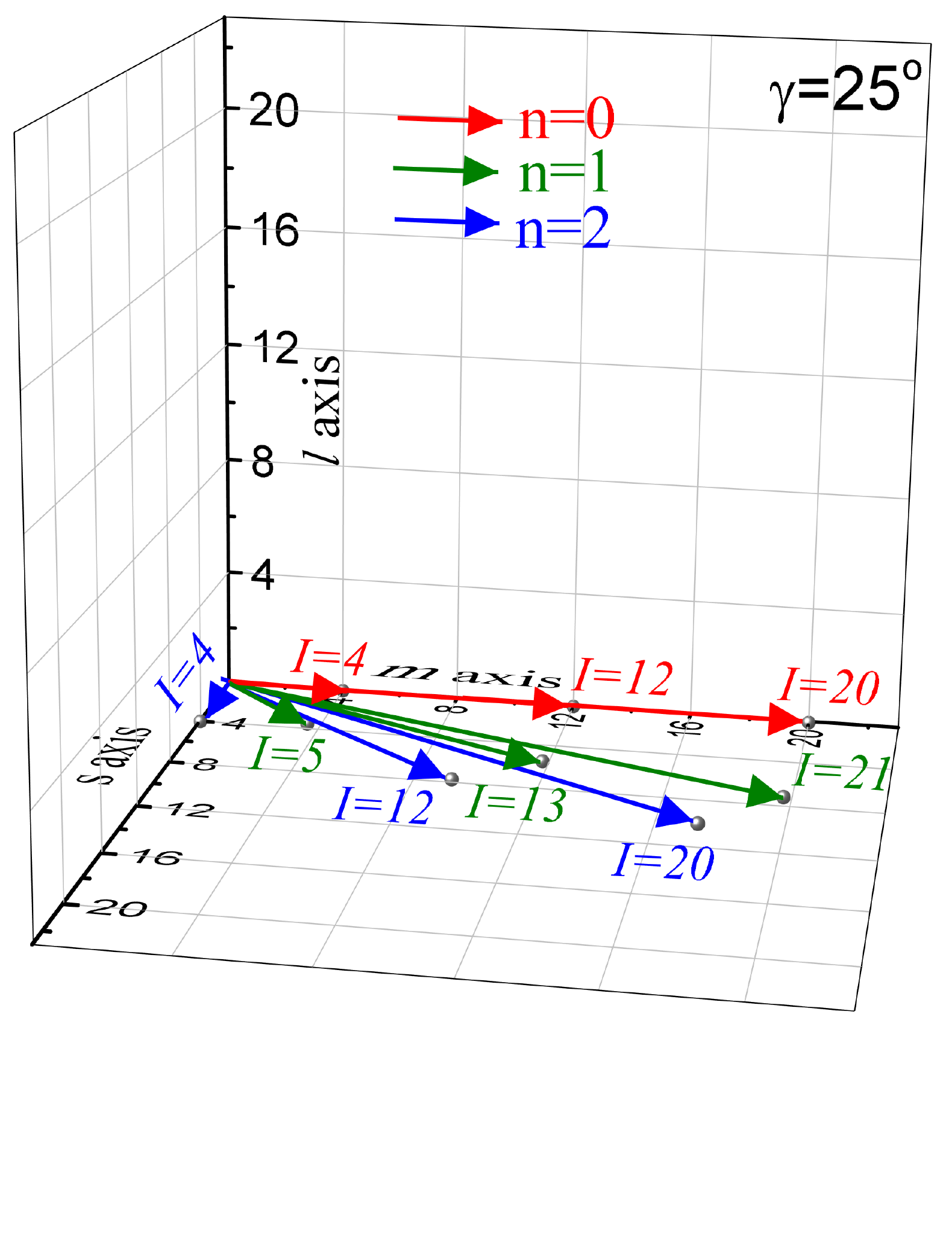}\\
\caption{\label{figAplotRu110}  Azimuthal plots for states $(n,I)$,
in which $I=4$, 5, 12, 13, 20, $21\hbar$ with $n=0$, 1, $2$, and the
schematic illustration of the evolution of angular momenta for the
wobbling band. Results for $\gamma=30^{\circ}$ and $25^{\circ}$ are
shown in upper and lower panels, respectively. }
\end{figure*}

\subsection{Two types of angular momentum geometries}

In this work, we want to illustrate the angular momentum geometry of
the wobbling motion by a probability density profile on the $(\theta, \varphi)$
unit sphere, called azimuthal plot~\cite{CFQ17, CQB18, Streck18}.
Here, $(\theta, \varphi)$ are the orientation angles of the angular momentum
vector $\textit{\textbf{I}}$ (expectation value with $M = I$ ) with
respect to the intrinsic frame. The polar angle $\theta$ is the
angle between $\textit{\textbf{I}}$ and the $l$-axis, whereas the
azimuthal angle $\varphi$ is the angle between the projection of $\textit{\textbf{I}}$ on
the $m$-$s$ plane and the $m$-axis. The profile
can be obtained by relating the orientation angles $(\theta, \varphi)$
to the Euler angles $(\psi, \theta, \pi-\varphi)$, where the $z$ axis
in the laboratory frame is chosen along $\textit{\textbf{I}}$. The
profile is calculated as~\cite{Streck18,CQB18}
\begin{align}
&\quad \mathcal{P}^{(\nu)}(\theta, \varphi)\notag\\
&=\langle I, \theta \varphi \mid I I \nu\rangle^{2}\notag\\
&=\sum_{K K^{\prime}} D_{K I}^{I *}(\theta, \varphi, 0)
C_{IK}^{(\nu)}C_{IK'}^{(\nu)} D_{K^{\prime} I}^{I}(\theta, \varphi, 0),
\end{align}
where $C^{\nu}_{IK}$ are the expansion coefficients in Eq.~(\ref{eqTRMwave}).

In Fig.~\ref{figAplot}, the obtained profiles ${\cal P} (\theta,
\varphi)$ are shown for the ground state and all of the wobbling
excited states at spin 12$\hbar$. To visualize the results of Fig.~\ref{figAplot},
we show the schematic of angular momenta geometry in Fig.~\ref{figAplot2}.
Note that the orientation of the angular momentum vector in this figure
just corresponds to the position of the maxima of ${\cal P} (\theta, \varphi)$.
We choose three results for $\gamma=30^{\circ}, 29^{\circ}$ and
$25^{\circ}$, whose ratio of ${\cal J}_m:{\cal J}_s:{\cal J}_l$
are  $4:1:1$, $4.3:1.1:1$ and $5.6: 1.8 : 1$, respectively.

\begin{figure*}[ht!]
\subfigure[~TRM with
$\gamma=30^{\circ}$]{\includegraphics[width=4.8cm]{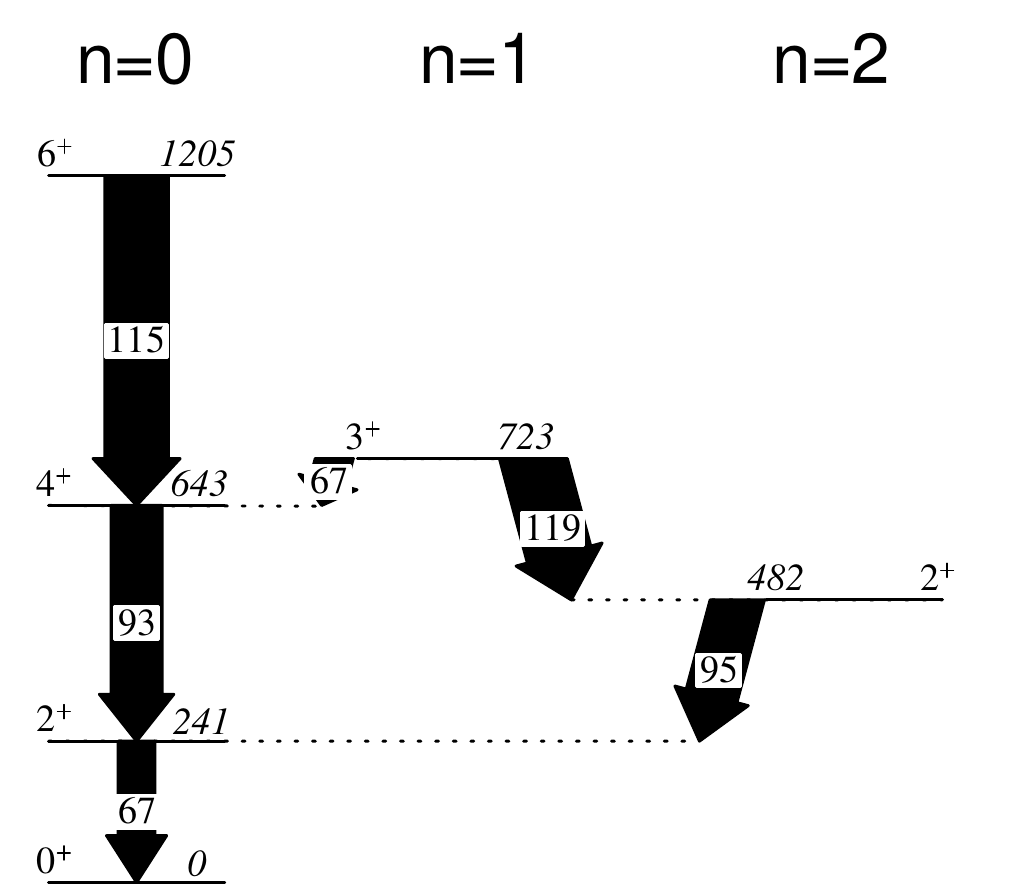}}
\subfigure[~Experiment of $^{110}$Ru
]{\includegraphics[width=5cm]{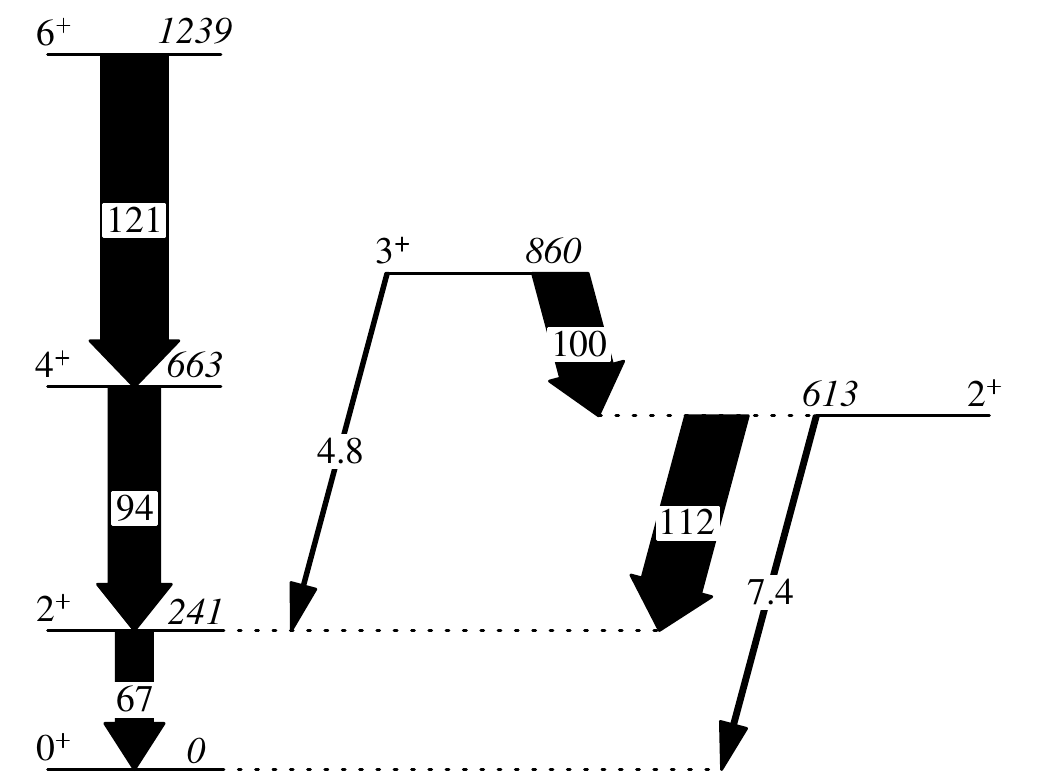}} \subfigure[~TRM with
$\gamma=25^{\circ}$]{\includegraphics[width=5.15cm]{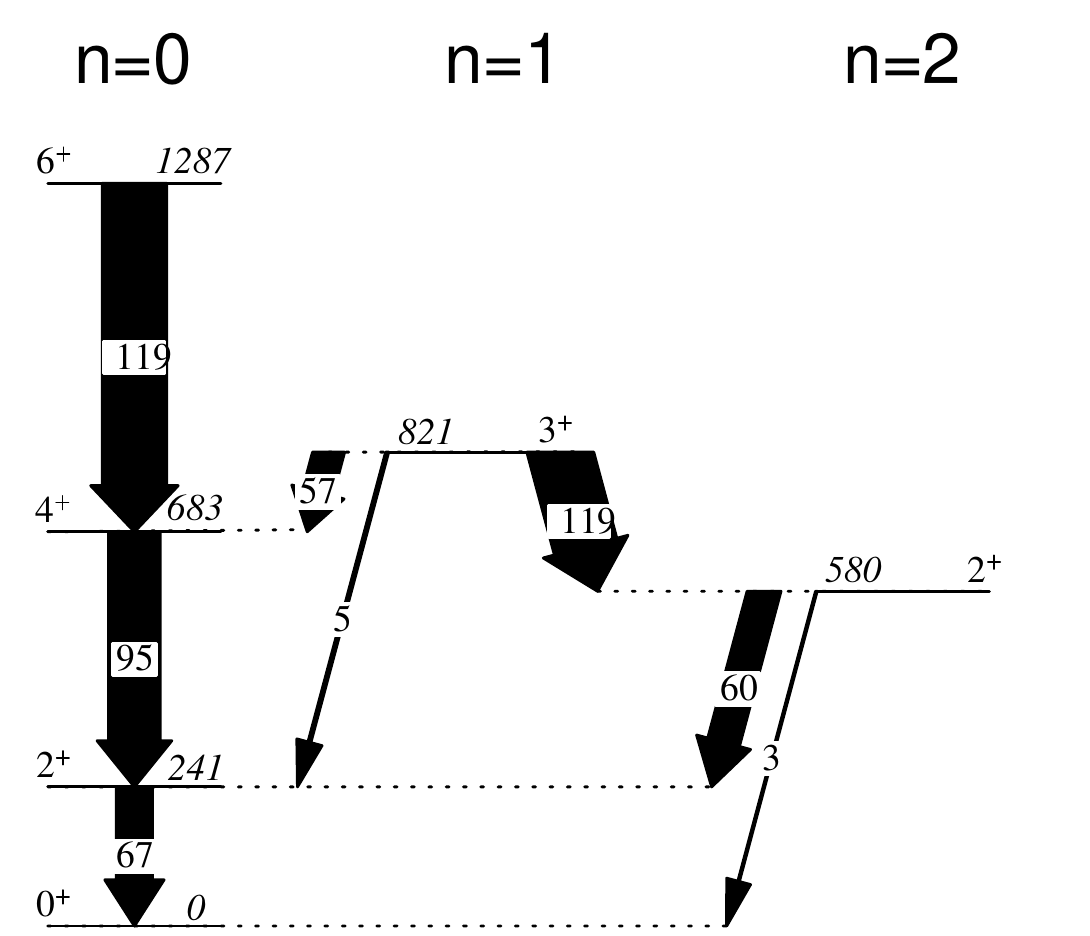}}
\caption{\label{figRu110} Comparisons between experimental level
scheme of $^{110}$Ru~\cite{Doherty17} and TRM results with $\gamma=30^{\circ}$
and $25^{\circ}$. The excitation energies (in keV) and spin-parity
values are given above the states. The widths and labels of the
arrows represent the reduced $E2$ transition
probabilities in W.u.}
\end{figure*}

One observes that the maximum of ${\cal P} (\theta, \varphi)$ is
always located at $\theta= 90^\circ, \varphi= 0^\circ$ for the
ground state, which means along the $m$-axis. The excited
states exhibit different features for different $\gamma$
values.

For $\gamma=25^\circ$ in the right panels of Fig.~\ref{figAplot},
${\cal P} (\theta, \varphi)$ show a very clear evolution of the
angular momentum as the increase in the phonon number: $m$-axis
$(n=0)$ $\longrightarrow $ $m$-$s$ plane $(n=2, 4)$
$\longrightarrow $  $s$-axis $(n=6)$ $\longrightarrow $ $s$-$l$
plane $(n=8, 10)$ $\longrightarrow $  $l$-axis $(n=12)$. This
process are also shown in Fig.~\ref{figAplot2}.
In some previous discussions, e.g., Ref.~\cite{Frauendorf14, CQB15},
similar picture was mentioned based on the case of ${\cal J}_m:{\cal
J}_s:{\cal J}_l=6: 2: 1$.

The picture of $\gamma=30^\circ$ is different from that of $25^\circ$.
The ${\cal P} (\theta, \varphi)$ is cylindrical symmetry with
$(\theta=90^\circ, \varphi=0^\circ)$ since $\mathcal{J}_s=\mathcal{J}_l$.
One notes that since the length of $s$- and $l$- axis are different,
such precessional motion of rotational axis with respect to $m$-axis
still makes sense. For $n=2, 4, 6$ states, the largest probability of
angular momentum has the radius of the circle about $30^\circ$, $45^\circ$
and $60^\circ$, respectively, indicating the amplitude of fluctuation of
the rotation axis is getting larger. The orientation of angular
momentum with the largest probability  are also shown in dotted
line in Fig.~\ref{figAplot2}.

The results of $\gamma=29^\circ$ is mixture of the character between
$\gamma=30^\circ$ and $25^\circ$ cases. $n=2$ state of $29^\circ$ is
close to the case of $30^\circ$, while $n=12$ state of $29^\circ$
is close to the case of $25^\circ$.

Furthermore, the azimuthal plots and the schematic of angular momenta
for the band of $n=0,1,2$ with the increase of spin are shown in
Fig.~\ref{figAplotRu110}. One observes that the angle between angular
momentum and $m$-axis decreases as the spin increasing, for both
$n=1$ and $n=2$ wobbling bands, which is consistent with the HA
formula with the precession amplitude $\sqrt{n/I}$.

\subsection{Comparison with the recent data of $^{110}$Ru}

Recently, a multi-step Coulomb excitation measurement was carried
out for $^{110}$Ru isotope~\cite{Doherty17}. The experimental data
of $^{110}$Ru are shown in Fig.~\ref{figRu110}(b), where the
excitation energies (in keV) and spin-parity values are given above
the states. The widths and labels of the arrows represent the
measured reduced $E2$ transition probabilities in
W.u.. It should be noted the $2^+_{2}$ and
$3^+_{1}$ are considered as one band in the Ref.~\cite{Doherty17},
while we separate them in the present level schemes.
Ref.~\cite{Doherty17} pointed out that the data provides direct
evidence of relatively rigid triaxial deformation near the ground
state.

In Fig.~\ref{figRu110}(a) and (c), we show the results calculated by
TRM with $\gamma=30^{\circ}$ and $25^{\circ}$. The adopted parameter
of MoI ${\cal J}_0$ ($\sim24~\hbar^2$MeV$^{-1}$) and quadruple
moment $Q$ ($\sim3.3$~$e$b) are adjusted for the energy  and
$B(E2)$ value  of $2_{1}^{+}$ state.

The calculated results are in agreement with the experimental data
qualitatively. As mentioned and emphasized in Ref.~\cite{Doherty17}, the
relatively large $2^+_2 \rightarrow 2^+_1$ and small $2^+_2 \rightarrow 0^+_1$
matrix elements, are strong indications of triaxial deformation. These
experimental characteristics are reproduced by the present calculations.
In addition, the large $3^+_1 \rightarrow 2^+_2$ and small $3^+_1 \rightarrow 2^+_1$
matrix elements in experiment are reproduced by TRM. Based on this,
3$_1^+$ and 2$_2^+$ for the $^{110}$Ru might be suggested as the bandhead
of the one- and two-phonon wobbling bands.

\section{Summary}

The influence of triaxial parameter $\gamma$ on the wobbling
excitation in even-even nuclei are investigated using the TRM
with the hydrodynamical MoIs. We suggest that the probability
of the $K_m=I-n$ component larger than 50\% might be a
reasonable criteria to judge the quality HA. Based on this
criteria and the characteristic of the energy spectra and
electric quadrupole transition probabilities,
wobbling motion in even-even nuclei could be realized well for
the states with small $n$ phonon number when $\gamma$ changing from
$\sim 25^{\circ}$ to $\sim 35^{\circ}$.

The above condition for the restriction of $\gamma$ value is a
relatively strict condition and might be difficult to achieve
in realistic nuclei, which might be one of reasons for
wobbling bands of purely collective were difficult to be
observed in experiment. A recent data from coulomb excitation
experiment, namely 3$_1^+$ and 2$_2^+$ for the $^{110}$Ru are
studied and might be suggested as the bandhead of the
candidate one- and two- phonon wobbling bands.

From azimuthal plot, the angular momentum geometry in the wobbling
excitation has two types due to the different MoI: one is exhibited
in the case of $\gamma\sim 30^\circ$ and the other one in $\gamma$
deviating from $30^\circ$. In a wobbling band with certain phonon
number, the angle between angular momentum and $m$-axis exhibits a
decreasing trend with respect to spin.

\begin{acknowledgments}
We are grateful to J. Meng and  P. W. Zhao  for helpful discussions.
This work is supported in parts by National Natural Science Foundation of
China (NSFC) under Grants No.~11675094 and 11622540, the Deutsche
Forschungsgemeinschaft (DFG) and the NSFC through funds provided to
the Sino-German CRC110 ``Symmetries and the Emergence of Structure
in QCD'' (DFG Grant No. TRR110 and NSFC Grant No. 11621131001).
\end{acknowledgments}

\end{CJK*}
\end{document}